\input harvmac
\let\includefigures=\iftrue
\let\useblackboard==\iftrue
\newfam\black

\includefigures
\message{If you do not have epsf.tex (to include figures),}
\message{change the option at the top of the tex file.}
\input epsf
\def\figin{\epsfcheck\figin}\def\figins{\epsfcheck\figins}
\def\epsfcheck{\ifx\epsfbox\UnDeFiNeD
\message{(NO epsf.tex, FIGURES WILL BE IGNORED)}
\gdef\figin##1{\vskip2in}\gdef\figins##1{\hskip.5in}
\else\message{(FIGURES WILL BE INCLUDED)}%
\gdef\figin##1{##1}\gdef\figins##1{##1}\fi}
\def\DefWarn#1{}
\def\figinsert{\goodbreak\midinsert}
\def\ifig#1#2#3{\DefWarn#1\xdef#1{fig.~\the\figno}
\writedef{#1\leftbracket fig.\noexpand~\the\figno}%
\figinsert\figin{\centerline{#3}}\medskip\centerline{\vbox{
\baselineskip12pt\advance\hsize by -1truein
\noindent\footnotefont{\bf Fig.~\the\figno:} #2}}
\endinsert\global\advance\figno by1}
\else
\def\ifig#1#2#3{\xdef#1{fig.~\the\figno}
\writedef{#1\leftbracket fig.\noexpand~\the\figno}%
\global\advance\figno by1} \fi

\def\id{{1 \kern-.28em {\rm l}}}

\def\K3{{\bf K3}}
\def\journal#1&#2(#3){\unskip, \sl #1\ \bf #2 \rm(19#3) }
\def\andjournal#1&#2(#3){\sl #1~\bf #2 \rm (19#3) }

\def\bar{\overline}
\def\hat{\widehat}
\def\ie{{\it i.e.}}

\def\tilde{\widetilde}

\def\frac#1#2{{#1\over#2}}

\def\inbar{\,\vrule height1.5ex width.4pt depth0pt}
\def\IC{\relax\hbox{$\inbar\kern-.3em{\rm C}$}}
\def\IR{\relax{\rm I\kern-.18em R}}
\def\IP{\relax{\rm I\kern-.18em P}}

%
%

%
\catcode`\@=11
\def\slash#1{\mathord{\mathpalette\c@ncel{#1}}}
\overfullrule=0pt

\def\underrel#1\over#2{\mathrel{\mathop{\kern\z@#1}\limits_{#2}}}

\catcode`\@=12


%

\def\tr{{\rm tr}}

\def \sinh{{\rm sinh}}
\def \cosh{{\rm cosh}}

\def\exp{{\rm exp}}


\def\ie{{\it i.e.}}


\lref\Zamolodchikov{
A.~B.~Zamolodchikov,
``Expectation value of composite field T anti-T in two-dimensional quantum field theory,''
[arXiv:hep-th/0401146 [hep-th]].
}

\lref\Smirnov{
F.~A.~Smirnov and A.~B.~Zamolodchikov,
``On space of integrable quantum field theories,''
Nucl. Phys. B {\bf 915}, 363-383 (2017)
doi:10.1016/j.nuclphysb.2016.12.014
[arXiv:1608.05499 [hep-th]].
}

\lref\Cavaglia{
A.~Cavagli\`a, S.~Negro, I.~M.~Sz\'ecs\'enyi and R.~Tateo,
``$T \bar{T}$-deformed 2D Quantum Field Theories,''
JHEP {\bf 10}, 112 (2016)
doi:10.1007/JHEP10(2016)112
[arXiv:1608.05534 [hep-th]].
}

\lref\Hassan{
S.~F.~Hassan and A.~Sen,
``Marginal deformations of WZNW and coset models from O(d,d) transformation,''
Nucl. Phys. B {\bf 405}, 143-165 (1993)
doi:10.1016/0550-3213(93)90429-S
[arXiv:hep-th/9210121 [hep-th]].
}

\lref\Boyda{
E.~K.~Boyda, S.~Ganguli, P.~Horava and U.~Varadarajan,
``Holographic protection of chronology in universes of the Godel type,''
Phys. Rev. D {\bf 67}, 106003 (2003)
doi:10.1103/PhysRevD.67.106003
[arXiv:hep-th/0212087 [hep-th]].
}

\lref\Gauntlett{
J.~P.~Gauntlett, J.~B.~Gutowski, C.~M.~Hull, S.~Pakis and H.~S.~Reall,
``All supersymmetric solutions of minimal supergravity in five- dimensions,''
Class. Quant. Grav. {\bf 20}, 4587-4634 (2003)
doi:10.1088/0264-9381/20/21/005
[arXiv:hep-th/0209114 [hep-th]].
}

\lref\Brace{
D.~Brace, C.~A.~R.~Herdeiro and S.~Hirano,
``Classical and quantum strings in compactified pp waves and Godel type universes,''
Phys. Rev. D {\bf 69}, 066010 (2004)
doi:10.1103/PhysRevD.69.066010
[arXiv:hep-th/0307265 [hep-th]].
}

\lref\Harmark{
T.~Harmark and T.~Takayanagi,
``Supersymmetric Godel universes in string theory,''
Nucl. Phys. B {\bf 662}, 3-39 (2003)
doi:10.1016/S0550-3213(03)00349-3
[arXiv:hep-th/0301206 [hep-th]].
}

\lref\Berenstein{
D.~E.~Berenstein, J.~M.~Maldacena and H.~S.~Nastase,
``Strings in flat space and pp waves from N=4 superYang-Mills,''
JHEP {\bf 04}, 013 (2002)
doi:10.1088/1126-6708/2002/04/013
[arXiv:hep-th/0202021 [hep-th]].
}

\lref\Araujo{
T.~Araujo, E.~\'O.~Colg\'ain, Y.~Sakatani, M.~M.~Sheikh-Jabbari and H.~Yavartanoo,
``Holographic integration of $T \bar{T}$ \& $J \bar{T}$ via $O(d,d)$,''
JHEP {\bf 03}, 168 (2019)
doi:10.1007/JHEP03(2019)168
[arXiv:1811.03050 [hep-th]].
}

\lref\ApoloTT{
L.~Apolo, S.~Detournay and W.~Song,
``TsT, $T\bar{T}$ and black strings,''
JHEP {\bf 06}, 109 (2020)
doi:10.1007/JHEP06(2020)109
[arXiv:1911.12359 [hep-th]].
}

\lref\Godel{
K.~Godel,
``An Example of a new type of cosmological solutions of Einstein's field equations of graviation,''
Rev. Mod. Phys. {\bf 21}, 447-450 (1949)
doi:10.1103/RevModPhys.21.447
}

\lref\Lunin{
O.~Lunin and J.~M.~Maldacena,
``Deforming field theories with U(1) x U(1) global symmetry and their gravity duals,''
JHEP {\bf 05}, 033 (2005)
doi:10.1088/1126-6708/2005/05/033
[arXiv:hep-th/0502086 [hep-th]].
}

\lref\Asrat{
M.~Asrat,
``$T{\bar T}$ and Holography,''
doi:10.6082/uchicago.3365
[arXiv:2112.02596 [hep-th]].
}

\lref\McGough{
L.~McGough, M.~Mezei and H.~Verlinde,
``Moving the CFT into the bulk with $ T\overline{T} $,''
JHEP {\bf 04}, 010 (2018)
doi:10.1007/JHEP04(2018)010
[arXiv:1611.03470 [hep-th]].
}

\lref\Berkovits{
N.~Berkovits, C.~Vafa and E.~Witten,
``Conformal field theory of AdS background with Ramond-Ramond flux,''
JHEP {\bf 03}, 018 (1999)
doi:10.1088/1126-6708/1999/03/018
[arXiv:hep-th/9902098 [hep-th]].
}

\lref\Gotz{
G.~Gotz, T.~Quella and V.~Schomerus,
``The WZNW model on PSU(1,1|2),''
JHEP {\bf 03}, 003 (2007)
doi:10.1088/1126-6708/2007/03/003
[arXiv:hep-th/0610070 [hep-th]].
}

\lref\Buser{
M.~Buser, E.~Kajari and W.~P.~Schleich,
``Visualization of the G\"odel universe,''
New J. Phys. {\bf 15}, 013063 (2013)
doi:10.1088/1367-2630/15/1/013063
[arXiv:1303.4651 [gr-qc]].
}

\lref\Buscher{
T.~H.~Buscher,
``A Symmetry of the String Background Field Equations,''
Phys. Lett. B {\bf 194}, 59-62 (1987)
doi:10.1016/0370-2693(87)90769-6
}

\lref\BuscherB{
T.~H.~Buscher,
``Path Integral Derivation of Quantum Duality in Nonlinear Sigma Models,''
Phys. Lett. B {\bf 201}, 466-472 (1988)
doi:10.1016/0370-2693(88)90602-8
}

\lref\Rocek{
M.~Rocek and E.~P.~Verlinde,
``Duality, quotients, and currents,''
Nucl. Phys. B {\bf 373}, 630-646 (1992)
doi:10.1016/0550-3213(92)90269-H
[arXiv:hep-th/9110053 [hep-th]].
}

\lref\Alvarez{
E.~Alvarez, L.~Alvarez-Gaume, J.~L.~F.~Barbon and Y.~Lozano,
``Some global aspects of duality in string theory,''
Nucl. Phys. B {\bf 415}, 71-100 (1994)
doi:10.1016/0550-3213(94)90067-1
[arXiv:hep-th/9309039 [hep-th]].
}

\lref\Hawking{
S.~W.~Hawking and G.~F.~R.~Ellis,
``The Large Scale Structure of Space-Time,''
Cambridge University Press, 2023,
ISBN 978-1-00-925316-1, 978-1-00-925315-4, 978-0-521-20016-5, 978-0-521-09906-6, 978-0-511-82630-6, 978-0-521-09906-6
doi:10.1017/9781009253161
}

\lref\Osten{
D.~Osten and S.~J.~van Tongeren,
``Abelian Yang-Baxter deformations and TsT transformations,''
Nucl. Phys. B {\bf 915}, 184-205 (2017)
doi:10.1016/j.nuclphysb.2016.12.007
[arXiv:1608.08504 [hep-th]].
}

\lref\BerkovitsB{
N.~Berkovits and J.~Maldacena,
``Fermionic T-Duality, Dual Superconformal Symmetry, and the Amplitude/Wilson Loop Connection,''
JHEP {\bf 09}, 062 (2008)
doi:10.1088/1126-6708/2008/09/062
[arXiv:0807.3196 [hep-th]].
}

\lref\Bekenstein{
J.~D.~Bekenstein,
``Black holes and entropy,''
Phys. Rev. D {\bf 7}, 2333-2346 (1973)
doi:10.1103/PhysRevD.7.2333
}

\lref\Wald{
R.~M.~Wald,
``Black hole entropy is the Noether charge,''
Phys. Rev. D {\bf 48}, no.8, R3427-R3431 (1993)
doi:10.1103/PhysRevD.48.R3427
[arXiv:gr-qc/9307038 [gr-qc]].
}

\lref\Sfondrini{
A.~Sfondrini and S.~J.~van Tongeren,
``$T\bar{T}$ deformations as $TsT$ transformations,''
Phys. Rev. D {\bf 101}, no.6, 066022 (2020)
doi:10.1103/PhysRevD.101.066022
[arXiv:1908.09299 [hep-th]].
}

\lref\Penrose{
R.~Penrose,
``Any Space-Time has a Plane Wave as a Limit,''
Springer Netherlands, Dordrecht, p. 271--27, ISBN 978-94-010-1508-0 (1976)
}

\lref\Gueven{
R.~Gueven,
``Plane wave limits and T duality,''
Phys. Lett. B {\bf 482}, 255-263 (2000)
doi:10.1016/S0370-2693(00)00517-7
[arXiv:hep-th/0005061 [hep-th]].
}

\lref\Freedman{
D.~Z.~Freedman, S.~S.~Gubser, K.~Pilch and N.~P.~Warner,
``Renormalization group flows from holography supersymmetry and a c theorem,''
Adv. Theor. Math. Phys. {\bf 3}, 363-417 (1999)
doi:10.4310/ATMP.1999.v3.n2.a7
[arXiv:hep-th/9904017 [hep-th]].
}

\lref\Girardello{
L.~Girardello, M.~Petrini, M.~Porrati and A.~Zaffaroni,
``Novel local CFT and exact results on perturbations of N=4 superYang Mills from AdS dynamics,''
JHEP {\bf 12}, 022 (1998)
doi:10.1088/1126-6708/1998/12/022
[arXiv:hep-th/9810126 [hep-th]].
}

\lref\Gubser{
S.~S.~Gubser,
``Curvature singularities: The Good, the bad, and the naked,''
Adv. Theor. Math. Phys. {\bf 4}, 679-745 (2000)
doi:10.4310/ATMP.2000.v4.n3.a6
[arXiv:hep-th/0002160 [hep-th]].
}

\lref\Dubovsky{
S.~Dubovsky, T.~Gregoire, A.~Nicolis and R.~Rattazzi,
``Null energy condition and superluminal propagation,''
JHEP {\bf 03}, 025 (2006)
doi:10.1088/1126-6708/2006/03/025
[arXiv:hep-th/0512260 [hep-th]].
}

\lref\Casimir{
H.~B.~G.~Casimir,
``On the Attraction Between Two Perfectly Conducting Plates,''
Indag. Math. {\bf 10}, 261-263 (1948)
}

\lref\Aharony{
O.~Aharony, S.~Datta, A.~Giveon, Y.~Jiang and D.~Kutasov,
``Modular invariance and uniqueness of $T\bar{T}$ deformed CFT,''
JHEP {\bf 01}, 086 (2019)
doi:10.1007/JHEP01(2019)086
[arXiv:1808.02492 [hep-th]].
}

\lref\AsratT{
M.~Asrat,
``KdV charges and the generalized torus partition sum in $T \bar T$ deformation,''
Nucl. Phys. B {\bf 958}, 115119 (2020)
doi:10.1016/j.nuclphysb.2020.115119
[arXiv:2002.04824 [hep-th]].
}

\lref\Chang{
C.~K.~Chang, C.~Ferko and S.~Sethi,
``Supersymmetry and $ T\overline{T} $ deformations,''
JHEP {\bf 04}, 131 (2019)
doi:10.1007/JHEP04(2019)131
[arXiv:1811.01895 [hep-th]].
}

\lref\Drukker{
N.~Drukker, B.~Fiol and J.~Simon,
``Godel's universe in a supertube shroud,''
Phys. Rev. Lett. {\bf 91}, 231601 (2003)
doi:10.1103/PhysRevLett.91.231601
[arXiv:hep-th/0306057 [hep-th]].
}

\lref\Gimon{
E.~G.~Gimon and P.~Horava,
``Over-rotating black holes, Godel holography and the hypertube,''
[arXiv:hep-th/0405019 [hep-th]].
}

\lref\Baggio{
M.~Baggio, A.~Sfondrini, G.~Tartaglino-Mazzucchelli and H.~Walsh,
``On $ T\overline{T} $ deformations and supersymmetry,''
JHEP {\bf 06}, 063 (2019)
doi:10.1007/JHEP06(2019)063
[arXiv:1811.00533 [hep-th]].
}

\lref\Forste{
S.~Forste,
``A Truly marginal deformation of SL(2, R) in a null direction,''
Phys. Lett. B {\bf 338}, 36-39 (1994)
doi:10.1016/0370-2693(94)91340-4
[arXiv:hep-th/9407198 [hep-th]].
}

\lref\IsraelM{
D.~Israel, C.~Kounnas and M.~P.~Petropoulos,
``Superstrings on NS5 backgrounds, deformed AdS(3) and holography,''
JHEP {\bf 10}, 028 (2003)
doi:10.1088/1126-6708/2003/10/028
[arXiv:hep-th/0306053 [hep-th]].
}

\lref\IsraelMM{
D.~Israel,
``Quantization of heterotic strings in a Godel / anti-de Sitter space-time and chronology protection,''
JHEP {\bf 01}, 042 (2004)
doi:10.1088/1126-6708/2004/01/042
[arXiv:hep-th/0310158 [hep-th]].
}

\lref\KiritsisM{
E.~Kiritsis,
``Exact duality symmetries in CFT and string theory,''
Nucl. Phys. B {\bf 405}, 109-142 (1993)
doi:10.1016/0550-3213(93)90428-R
[arXiv:hep-th/9302033 [hep-th]].
}

\lref\HorowitzM{
G.~T.~Horowitz and A.~A.~Tseytlin,
``A New class of exact solutions in string theory,''
Phys. Rev. D {\bf 51}, 2896-2917 (1995)
doi:10.1103/PhysRevD.51.2896
[arXiv:hep-th/9409021 [hep-th]].
}

\lref\Sadri{
D.~Sadri and M.~M.~Sheikh-Jabbari,
``String theory on parallelizable pp waves,''
JHEP {\bf 06}, 005 (2003)
doi:10.1088/1126-6708/2003/06/005
[arXiv:hep-th/0304169 [hep-th]].
}

\lref\Frolov{
S.~Frolov,
``Lax pair for strings in Lunin-Maldacena background,''
JHEP {\bf 05}, 069 (2005)
doi:10.1088/1126-6708/2005/05/069
[arXiv:hep-th/0503201 [hep-th]].
}

\lref\Mateos{
D.~Mateos and P.~K.~Townsend,
``Supertubes,''
Phys. Rev. Lett. {\bf 87}, 011602 (2001)
doi:10.1103/PhysRevLett.87.011602
[arXiv:hep-th/0103030 [hep-th]].
}

\lref\Bousso{
R.~Bousso,
``The Holographic principle for general backgrounds,''
Class. Quant. Grav. {\bf 17}, 997-1005 (2000)
doi:10.1088/0264-9381/17/5/309
[arXiv:hep-th/9911002 [hep-th]].
}

\lref\HawkingM{
S.~W.~Hawking,
``The Chronology protection conjecture,''
Phys. Rev. D {\bf 46}, 603-611 (1992)
doi:10.1103/PhysRevD.46.603
}

\lref\Johnson{
C.~V.~Johnson, A.~W.~Peet and J.~Polchinski,
``Gauge theory and the excision of repulson singularities,''
Phys. Rev. D {\bf 61}, 086001 (2000)
doi:10.1103/PhysRevD.61.086001
[arXiv:hep-th/9911161 [hep-th]].
}

\lref\SomMM{
M.~M.~Som and  A.~K.~Raychaudhuri,
``Cylindrically symmetric charged dust distributions in rigid rotation in general relativity,"
Proc. R. Soc. Lond. A {\bf 304} 81-86 (1968)
doi:10.1098/rspa.1968.0073
}

\lref\Brecher{
D.~Brecher, P.~A.~DeBoer, D.~C.~Page and M.~Rozali,
``Closed time - like curves and holography in compact plane waves,''
JHEP {\bf 10}, 031 (2003)
doi:10.1088/1126-6708/2003/10/031
[arXiv:hep-th/0306190 [hep-th]].
}

\lref\WilsonMM{
K.~G.~Wilson and W.~Zimmermann,
``Operator product expansions and composite field operators in the general framework of quantum field theory,''
Commun. Math. Phys. {\bf 24} (1972), 87-106
doi:10.1007/BF01878448
}

\lref\WittenM{
E.~Witten,
``Multitrace operators, boundary conditions, and AdS / CFT correspondence,''
[arXiv:hep-th/0112258 [hep-th]].
}

\lref\AldayMM{
L.~F.~Alday, G.~Arutyunov and S.~Frolov,
``Green-Schwarz strings in TsT-transformed backgrounds,''
JHEP {\bf 06}, 018 (2006)
doi:10.1088/1126-6708/2006/06/018
[arXiv:hep-th/0512253 [hep-th]].
}

\lref\Lowenstein{
J.~H.~Lowenstein,
``Normal products in the thirring model,''
Commun. Math. Phys. {\bf 16}, 265-289 (1970)
doi:10.1007/BF01646535
}

\lref\ApoloMM{
L.~Apolo, P.~X.~Hao, W.~X.~Lai and W.~Song,
``Glue-on AdS holography for $ T\overline{T} $-deformed CFTs,''
JHEP {\bf 06}, 117 (2023)
doi:10.1007/JHEP06(2023)117
[arXiv:2303.04836 [hep-th]].
}

\lref\ManschotMM{
J.~Manschot and S.~Mondal,
``Supersymmetric black holes and $T{\bar T}$ deformation,''
Phys. Rev. D {\bf 107}, no.12, L121903 (2023)
doi:10.1103/PhysRevD.107.L121903
[arXiv:2207.01462 [hep-th]].
}

\lref\MaldacenaMM{J.~M.~Maldacena,
``The Large N limit of superconformal field theories and supergravity,''
Adv. Theor. Math. Phys. {\bf 2}, 231-252 (1998)
doi:10.4310/ATMP.1998.v2.n2.a1
[arXiv:hep-th/9711200 [hep-th]].
}

\lref\GubserMM{
S.~S.~Gubser, I.~R.~Klebanov and A.~M.~Polyakov,
``Gauge theory correlators from noncritical string theory,''
Phys. Lett. B {\bf 428}, 105-114 (1998)
doi:10.1016/S0370-2693(98)00377-3
[arXiv:hep-th/9802109 [hep-th]].
}

\lref\WittenMM{
E.~Witten,
``Anti-de Sitter space and holography,''
Adv. Theor. Math. Phys. {\bf 2}, 253-291 (1998)
doi:10.4310/ATMP.1998.v2.n2.a2
[arXiv:hep-th/9802150 [hep-th]].
}

\lref\MMRazamat{
O.~Aharony, Z.~Komargodski and S.~S.~Razamat,
``On the worldsheet theories of strings dual to free large N gauge theories,''
JHEP {\bf 05}, 016 (2006)
doi:10.1088/1126-6708/2006/05/016
[arXiv:hep-th/0602226 [hep-th]].
}

\lref\HerdeiroMM{
C.~A.~R.~Herdeiro,
``Spinning deformations of the D1 - D5 system and a geometric resolution of closed timelike curves,''
Nucl. Phys. B {\bf 665}, 189-210 (2003)
doi:10.1016/S0550-3213(03)00484-X
[arXiv:hep-th/0212002 [hep-th]].
}

\lref\KutasovME{
D.~Kutasov and N.~Seiberg,
``More comments on string theory on AdS(3),''
JHEP {\bf 04}, 008 (1999)
doi:10.1088/1126-6708/1999/04/008
[arXiv:hep-th/9903219 [hep-th]].
}

\lref\AharonyMMR{
O.~Aharony, M.~Berkooz and E.~Silverstein,
``Multiple trace operators and nonlocal string theories,''
JHEP {\bf 08}, 006 (2001)
doi:10.1088/1126-6708/2001/08/006
[arXiv:hep-th/0105309 [hep-th]].
}

\lref\BeckerME{
K.~Becker and M.~Becker,
``Interactions in the SL(2,IR) / U(1) black hole background,''
Nucl. Phys. B {\bf 418}, 206-230 (1994)
doi:10.1016/0550-3213(94)90245-3
[arXiv:hep-th/9310046 [hep-th]].
}

\lref\FDavidMM{
F.~David,
``Conformal Field Theories Coupled to 2D Gravity in the Conformal Gauge,''
Mod. Phys. Lett. A {\bf 3}, 1651 (1988)
doi:10.1142/S0217732388001975
}

\lref\ChaudhuriS{
S.~Chaudhuri and J.~A.~Schwartz,
``A Criterion for Integrably Marginal Operators,''
Phys. Lett. B {\bf 219}, 291-296 (1989)
doi:10.1016/0370-2693(89)90393-6
}

\lref\Courant{
T.~Courant,
``Dirac manifolds,''
Trans.Am.Math.Soc. {\bf 319} (1990) 631
doi:10.1090/S0002-9947-1990-0998124-1
}

\lref\HenneauxMMT{
M.~Henneaux, C.~Martinez, R.~Troncoso and J.~Zanelli,
``Asymptotically anti-de Sitter spacetimes and scalar fields with a logarithmic branch,''
Phys. Rev. D {\bf 70}, 044034 (2004)
doi:10.1103/PhysRevD.70.044034
[arXiv:hep-th/0404236 [hep-th]].
}

\lref\ReggeMMT{
T.~Regge and C.~Teitelboim,
``Role of Surface Integrals in the Hamiltonian Formulation of General Relativity,''
Annals Phys. {\bf 88}, 286 (1974)
doi:10.1016/0003-4916(74)90404-7
}

\lref\HenneauxMMTT{
M.~Henneaux, C.~Martinez, R.~Troncoso and J.~Zanelli,
``Asymptotic behavior and Hamiltonian analysis of anti-de Sitter gravity coupled to scalar fields,''
Annals Phys. {\bf 322}, 824-848 (2007)
doi:10.1016/j.aop.2006.05.002
[arXiv:hep-th/0603185 [hep-th]].
}

\lref\Sokoloff{
D.~D.~Sokoloff, A.~A.~Starobinskii, ``On the structure of curvature tensor on conical singularities,"
Dokl. Akad. Nauk SSSR, 1977, Volume {\bf 234}, Number 5, 1043-1046
}

\lref\Kowalski{
Kowalski, O.~, Szenthe, J~. ``On the Existence of Homogeneous Geodesics in Homogeneous Riemannian Manifolds,"
Geometriae Dedicata {\bf 81}, 209-214 (2000)
doi:10.1023/A:1005287907806
}

\lref\GauntlettNM{
J.~P.~Gauntlett, J.~B.~Gutowski, C.~M.~Hull, S.~Pakis and H.~S.~Reall,
``All supersymmetric solutions of minimal supergravity in five- dimensions,''
Class. Quant. Grav. {\bf 20}, 4587-4634 (2003)
doi:10.1088/0264-9381/20/21/005
[arXiv:hep-th/0209114 [hep-th]].
}

\lref\MaldacenaDLC{
J.~Maldacena, D.~Martelli and Y.~Tachikawa,
``Comments on string theory backgrounds with non-relativistic conformal symmetry,''
JHEP {\bf 10}, 072 (2008)
doi:10.1088/1126-6708/2008/10/072
[arXiv:0807.1100 [hep-th]].
}

\lref\CallanCGS{
C.~G.~Callan, Jr., E.~J.~Martinec, M.~J.~Perry and D.~Friedan,
``Strings in Background Fields,''
Nucl. Phys. B {\bf 262}, 593-609 (1985)
doi:10.1016/0550-3213(85)90506-1
}

\lref\DijkgraafMM{
R.~Dijkgraaf, H.~L.~Verlinde and E.~P.~Verlinde,
``String propagation in a black hole geometry,''
Nucl. Phys. B {\bf 371}, 269-314 (1992)
doi:10.1016/0550-3213(92)90237-6
}

\lref\GinspargM{
P.~H.~Ginsparg and F.~Quevedo,
``Strings on curved space-times: Black holes, torsion, and duality,''
Nucl. Phys. B {\bf 385}, 527-557 (1992)
doi:10.1016/0550-3213(92)90057-I
[arXiv:hep-th/9202092 [hep-th]].
}

\lref\SfetsosM{
K.~Sfetsos,
``Conformally exact results for SL(2,R) x SO(1,1)(d-2) / SO(1,1) coset models,''
Nucl. Phys. B {\bf 389}, 424-444 (1993)
doi:10.1016/0550-3213(93)90327-L
[arXiv:hep-th/9206048 [hep-th]].
}

\lref\Codina{
T.~Codina, O.~Hohm and B.~Zwiebach,
``On black hole singularity resolution in $D=2$ via duality-invariant $\alpha'$ corrections,''
[arXiv:2308.09743 [hep-th]].
}

\lref\WittenMMM{
E.~Witten,
``On string theory and black holes,''
Phys. Rev. D {\bf 44}, 314-324 (1991)
doi:10.1103/PhysRevD.44.314
}

\lref\Mandal{
G.~Mandal, A.~M.~Sengupta and S.~R.~Wadia,
``Classical solutions of two-dimensional string theory,''
Mod. Phys. Lett. A {\bf 6}, 1685-1692 (1991)
doi:10.1142/S0217732391001822
}

\lref\GiveonMMMMMM{
A.~Giveon,
``Target space duality and stringy black holes,''
Mod. Phys. Lett. A {\bf 6}, 2843-2854 (1991)
doi:10.1142/S0217732391003316
}

\lref\HullMMMM{
C.~M.~Hull,
``Timelike T duality, de Sitter space, large N gauge theories and topological field theory,''
JHEP {\bf 07}, 021 (1998)
doi:10.1088/1126-6708/1998/07/021
[arXiv:hep-th/9806146 [hep-th]].
}

\lref\AsratMHB{
M.~Asrat,
``Moving holographic boundaries,''
[arXiv:2305.15744 [hep-th]].
}

\lref\Giveon{
A.~Giveon, N.~Itzhaki and D.~Kutasov,
``$T{\bar T} $ and LST,''
JHEP {\bf 07}, 122 (2017)
doi:10.1007/JHEP07(2017)122
[arXiv:1701.05576 [hep-th]].
}

\lref\Floratos{
E.~G.~Floratos, D.~A.~Ross and C.~T.~Sachrajda,
``Higher Order Effects in Asymptotically Free Gauge Theories: The Anomalous Dimensions of Wilson Operators,''
Nucl. Phys. B {\bf 129}, 66-88 (1977)
[erratum: Nucl. Phys. B {\bf 139}, 545-546 (1978)]
doi:10.1016/0550-3213(77)90020-7
}

\lref\FloratosA{
E.~G.~Floratos, D.~A.~Ross and C.~T.~Sachrajda,
``Higher Order Effects in Asymptotically Free Gauge Theories. 2. Flavor Singlet Wilson Operators and Coefficient Functions,''
Nucl. Phys. B {\bf 152}, 493-520 (1979)
doi:10.1016/0550-3213(79)90094-4
}

\lref\Gonzalez{
A.~Gonzalez-Arroyo, C.~Lopez and F.~J.~Yndurain,
``Second Order Contributions to the Structure Functions in Deep Inelastic Scattering. 1. Theoretical Calculations,''
Nucl. Phys. B {\bf 153}, 161-186 (1979)
doi:10.1016/0550-3213(79)90596-0
}

\lref\GonzalezC{
A.~Gonzalez-Arroyo and C.~Lopez,
``Second Order Contributions to the Structure Functions in Deep Inelastic Scattering. 3. The Singlet Case,''
Nucl. Phys. B {\bf 166}, 429-459 (1980)
doi:10.1016/0550-3213(80)90207-2
}

\lref\Curci{
G.~Curci, W.~Furmanski and R.~Petronzio,
``Evolution of Parton Densities Beyond Leading Order: The Nonsinglet Case,''
Nucl. Phys. B {\bf 175}, 27-92 (1980)
doi:10.1016/0550-3213(80)90003-6
}

\lref\Parisi{
G.~Parisi,
``How to measure the dimension of the parton field,''
Nucl. Phys. B {\bf 59}, 641-646 (1973)
doi:10.1016/0550-3213(73)90666-4
}

\lref\Callan{
C.~G.~Callan, Jr. and D.~J.~Gross,
``Bjorken scaling in quantum field theory,''
Phys. Rev. D {\bf 8}, 4383-4394 (1973)
doi:10.1103/PhysRevD.8.4383
}

\lref\Gross{
D.~J.~Gross,
``Applications of the Renormalization Group to High-Energy Physics,''
Conf. Proc. C {\bf 7507281}, 141-250 (1975)
}

\lref\Vega{
H.~J.~de Vega and I.~L.~Egusquiza,
``Planetoid string solutions in (3+1) axisymmetric space-times,''
Phys. Rev. D {\bf 54}, 7513-7519 (1996)
doi:10.1103/PhysRevD.54.7513
[arXiv:hep-th/9607056 [hep-th]].
}

\lref\Gubser{
S.~S.~Gubser, I.~R.~Klebanov and A.~M.~Polyakov,
``A Semiclassical limit of the gauge / string correspondence,''
Nucl. Phys. B {\bf 636}, 99-114 (2002)
doi:10.1016/S0550-3213(02)00373-5
[arXiv:hep-th/0204051 [hep-th]].
}

\lref\Kruczenski{
M.~Kruczenski,
``Spiky strings and single trace operators in gauge theories,''
JHEP {\bf 08}, 014 (2005)
doi:10.1088/1126-6708/2005/08/014
[arXiv:hep-th/0410226 [hep-th]].
}

\lref\Burden{
C.~J.~Burden,
``Gravitational Radiation From a Particular Class of Cosmic Strings,''
Phys. Lett. B {\bf 164}, 277-281 (1985)
doi:10.1016/0370-2693(85)90326-0
}

\lref\Jevicki{
A.~Jevicki, K.~Jin, C.~Kalousios and A.~Volovich,
``Generating AdS String Solutions,''
JHEP {\bf 03}, 032 (2008)
doi:10.1088/1126-6708/2008/03/032
[arXiv:0712.1193 [hep-th]].
}

\lref\JevickiK{
A.~Jevicki and K.~Jin,
``Solitons and AdS String Solutions,''
Int. J. Mod. Phys. A {\bf 23}, 2289-2298 (2008)
doi:10.1142/S0217751X0804113X
[arXiv:0804.0412 [hep-th]].
}

\lref\DeVega{
H.~J.~De Vega and N.~G.~Sanchez,
``Exact integrability of strings in D-Dimensional De Sitter space-time,''
Phys. Rev. D {\bf 47}, 3394-3405 (1993)
doi:10.1103/PhysRevD.47.3394
}

\lref\Pohlmeyer{
K.~Pohlmeyer,
``Integrable Hamiltonian Systems and Interactions Through Quadratic Constraints,''
Commun. Math. Phys. {\bf 46}, 207-221 (1976)
doi:10.1007/BF01609119
}

\lref\LarsenA{
A.~L.~Larsen and N.~G.~Sanchez,
``Sinh-Gordon, cosh-Gordon and Liouville equations for strings and multistrings in constant curvature space-times,''
Phys. Rev. D {\bf 54}, 2801-2807 (1996)
doi:10.1103/PhysRevD.54.2801
[arXiv:hep-th/9603049 [hep-th]].
}

\lref\Korchemsky{
G.~P.~Korchemsky and G.~Marchesini,
``Structure function for large x and renormalization of Wilson loop,''
Nucl. Phys. B {\bf 406}, 225-258 (1993)
doi:10.1016/0550-3213(93)90167-N
[arXiv:hep-ph/9210281 [hep-ph]].
}

\lref\KorchemskyA{
G.~P.~Korchemsky,
``Asymptotics of the Altarelli-Parisi-Lipatov Evolution Kernels of Parton Distributions,''
Mod. Phys. Lett. A {\bf 4}, 1257-1276 (1989)
doi:10.1142/S0217732389001453
}

\lref\Alday{
L.~F.~Alday and J.~M.~Maldacena,
``Comments on operators with large spin,''
JHEP {\bf 11} (2007), 019
doi:10.1088/1126-6708/2007/11/019
[arXiv:0708.0672 [hep-th]].
}

\lref\AldayA{
L.~F.~Alday and A.~Zhiboedov,
``Conformal Bootstrap With Slightly Broken Higher Spin Symmetry,''
JHEP {\bf 06}, 091 (2016)
doi:10.1007/JHEP06(2016)091
[arXiv:1506.04659 [hep-th]].
}

\lref\Belitsky{
A.~V.~Belitsky, A.~S.~Gorsky and G.~P.~Korchemsky,
``Gauge / string duality for QCD conformal operators,''
Nucl. Phys. B {\bf 667}, 3-54 (2003)
doi:10.1016/S0550-3213(03)00542-X
[arXiv:hep-th/0304028 [hep-th]].
}

\lref\Callan{
C.~G.~Callan, Jr. and D.~J.~Gross,
``Bjorken scaling in quantum field theory,''
Phys. Rev. D {\bf 8}, 4383-4394 (1973)
doi:10.1103/PhysRevD.8.4383
}

\lref\Drukker{
N.~Drukker, D.~J.~Gross and H.~Ooguri,
``Wilson loops and minimal surfaces,''
Phys. Rev. D {\bf 60}, 125006 (1999)
doi:10.1103/PhysRevD.60.125006
[arXiv:hep-th/9904191 [hep-th]].
}

\lref\KruczenskiA{
M.~Kruczenski,
``A Note on twist two operators in N=4 SYM and Wilson loops in Minkowski signature,''
JHEP {\bf 12}, 024 (2002)
doi:10.1088/1126-6708/2002/12/024
[arXiv:hep-th/0210115 [hep-th]].
}

\lref\Basso{
B.~Basso, G.~P.~Korchemsky and J.~Kotanski,
``Cusp anomalous dimension in maximally supersymmetric Yang-Mills theory at strong coupling,''
Phys. Rev. Lett. {\bf 100}, 091601 (2008)
doi:10.1103/PhysRevLett.100.091601
[arXiv:0708.3933 [hep-th]].
}

\lref\MaldacenaA{
J.~M.~Maldacena,
``Wilson loops in large N field theories,''
Phys. Rev. Lett. {\bf 80}, 4859-4862 (1998)
doi:10.1103/PhysRevLett.80.4859
[arXiv:hep-th/9803002 [hep-th]].
}

\lref\Belitsky{
A.~V.~Belitsky, G.~P.~Korchemsky and R.~S.~Pasechnik,
``Fine structure of anomalous dimensions in N=4 super Yang-Mills theory,''
Nucl. Phys. B {\bf 809}, 244-278 (2009)
doi:10.1016/j.nuclphysb.2008.10.013
[arXiv:0806.3657 [hep-ph]].
}

\lref\BelitskyA{
A.~V.~Belitsky, A.~S.~Gorsky and G.~P.~Korchemsky,
``Logarithmic scaling in gauge/string correspondence,''
Nucl. Phys. B {\bf 748}, 24-59 (2006)
doi:10.1016/j.nuclphysb.2006.04.030
[arXiv:hep-th/0601112 [hep-th]].
}

\lref\BurdenA{
C.~J.~Burden and L.~J.~Tassie,
``ADDITIONAL RIGIDLY ROTATING SOLUTIONS IN THE STRING MODEL OF HADRONS,''
Austral. J. Phys. {\bf 37}, 1-7 (1984)
doi:10.1071/PH840001
}

\lref\MaldacenaW{
J.~M.~Maldacena,
``Wilson loops in large N field theories,''
Phys. Rev. Lett. {\bf 80}, 4859-4862 (1998)
doi:10.1103/PhysRevLett.80.4859
[arXiv:hep-th/9803002 [hep-th]].
}

\lref\KorchemskyW{
G.~P.~Korchemsky and A.~V.~Radyushkin,
``Renormalization of the Wilson Loops Beyond the Leading Order,''
Nucl. Phys. B {\bf 283}, 342-364 (1987)
doi:10.1016/0550-3213(87)90277-X
}
 
\lref\Meseret{
M.~Asrat,
``$T{\bar T}$ and Holography,''
doi:10.6082/uchicago.3365
[arXiv:2112.02596 [hep-th]].
}

\lref\KruczenskiRT{
M.~Kruczenski, R.~Roiban, A.~Tirziu and A.~A.~Tseytlin,
``Strong-coupling expansion of cusp anomaly and gluon amplitudes from quantum open strings in AdS(5) x S**5,''
Nucl. Phys. B {\bf 791}, 93-124 (2008)
doi:10.1016/j.nuclphysb.2007.09.005
[arXiv:0707.4254 [hep-th]].
}

\lref\Braga{
N.~R.~F.~Braga and E.~Iancu,
``Anomalous dimensions from rotating open strings in AdS/CFT,''
JHEP {\bf 08}, 104 (2014)
doi:10.1007/JHEP08(2014)104
[arXiv:1405.7388 [hep-th]].
}

\Title{
}
{\vbox{
\centerline{Rotating strings and anomalous dimensions } 
\bigskip
\centerline{ in Non-AdS holography }
}
}

\bigskip
\centerline{\it Meseret Asrat}
\smallskip
\centerline{International Center for Theoretical Sciences}
\centerline{Tata Institute of Fundamental Research
} \centerline{Bengaluru, KA 560089, India}

\smallskip

\vglue .3cm

\bigskip

\let\includefigures=\iftrue
\bigskip
\noindent

In this paper we consider certain rigidly rotating closed string configurations in an asymptotically non-AdS string background. The string background is a deformation of $AdS_3 \times {\cal M}_7$. It interpolates between $AdS_3 $ and asymptotically linear dilaton $\IR \times S^1 \times \IR $ spacetime (times the internal compact manifold ${\cal M}_7$). We compute the quantity $E - J$ (in the large $J$ limit) where $E$ is the energy and $J$ is the angular momentum of the spinning strings. In the two dimensional CFT dual to string theory on $AdS_3$ (times ${\cal M}_7$) it gives the anomalous dimensions of certain twist two and higher operators. We show in the deformed background that $E - J$ is bounded. At a special value of the deformation coupling we also show that for spinning closed strings containing $n > 2$ cusps or spikes both $E$ and $J$ are bounded. In the CFT dual to string theory on $AdS_3$ (times ${\cal M}_7$) the spinning cusped strings describe operators with twist $n$ larger than two. In general, at other values of the deformation coupling, we demonstrate that this feature is exhibited only by those cusped strings with $n > n_0$ where $n_0$ is determined only by the deformation coupling. We also give simple exact Regge relations between $E$ and $J$. We also study the closely related cusp anomalous dimension of a light-like Wilson loop. We show in perturbation that $E - J$ for a folded rodlike spinning closed string agrees qualitatively (but not quantitatively) with the cusp anomalous dimension. We comment on what $E - J$ measures away from the CFT along the deformation in the coupling space. In the long string sector the deformation is dual to a single trace $T{\bar T}$ deformed orbifold theory. We determine the associated deformed sinh-Gordon model that classically describes the (long) strings near the boundary. This provides an example of single trace $T{\bar T}$ deformation in non-orbifold theories.

\bigskip

\Date{04/23}


\newsec{Introduction}

We study a class of rotating closed string solutions in the string background obtained in \AsratMHB. The metric has the topology of ${\cal A}_3 \times S^3 \times {\cal X}_4$ where $S^3$ denotes a unit three-sphere and ${\cal X}_4$ is a compact four dimensional manifold. The simplest examples are the Calabi-Yau manifolds $T^4$ and $K_3$. The metric ${\cal A}_3$ interpolates between $AdS_3$ in the infrared and a (an asymptotically) linear dilaton spacetime $\IR \times S^1\times \IR$ in the ultraviolet. The worldsheet conformal field theory on ${\cal A}_3$ is an exact marginal current bilinear deformation of the worldsheet theory on $AdS_3$. In the boundary theory the deformation is equivalent to a deformation by an irrelevant operator of (left and right) dimensions $(2, 2)$ \Giveon. See also \AsratMHB\ and references therein. The resulting deformed theory is thus non-conformal and dual to string theory on ${\cal A}_3$ (times a compact internal manifold).

In the context of gauge/string equivalence, gauge theory states with large spin and lying on the leading Regge trajectory are described in the dual string theory by classical spinning strings \refs{\Vega\Gubser-\Kruczenski}. The leading Regge trajectory is consist of states with the lowest mass for each given spin. It is the (linear) curve that has the largest intercept with the spin axis. In the large spin limit the leading (one-loop) term of their anomalous dimensions is given (at weak coupling) by $\lambda\ln S$ where $S$ is the spin and $\lambda$ is the (relevant) 't Hooft coupling. The logarithmic scaling of anomalous dimensions is a universal large spin feature in all gauge theories and asymptotically free theories (including non-supersymmetric ones) \refs{\Parisi\Callan-\Gross,\ \BelitskyA}. Two and higher loop analyses in gauge theories indicate that in perturbation theory the anomalous dimensions should not grow (at each loop order) faster than $\ln S$ \refs{\Floratos\FloratosA-\Gonzalez}. This behavior of the anomalous dimensions for large spin is also generally believed to hold non-perturbatively \refs{\Gubser,\ \Korchemsky-\KorchemskyA}.

In this paper we consider a class of rotating closed strings in ${\cal A}_3$. In the undeformed two dimensional conformal field theory (CFT) dual to string theory on $AdS_3$ they describe twist two and higher operators. The twist of an operator is defined as the difference of its canonical scaling or mass dimension $\Delta$ and (Lorentz or conformal) spin $S$. For example, in four spacetime dimensions, the covariant derivative $\nabla_\mu$ has twist zero and the operators $\nabla^2 := \nabla_\mu\nabla^\mu$ and $\tr \left\{\phi\nabla_\mu\nabla_\nu\phi\right\}$ both have twist two where $\phi$ is a scalar and the curly brackets denote the symmetric traceless part. In deep inelastic scattering at high values of momentum transfer (and weak coupling) the dominant contributions to the structure (moment) functions come from single trace twist two operators. Thus, their anomalous dimensions are of particular interest in hadron physics \refs{\Gonzalez,\ \GonzalezC, \ \Curci}. An example of twist two and spin $n$ which is of particular interest in perturbative quantum chromodynamics (QCD) is the Wilson operator. It is schematically of the form $\left\{{\bar \psi} \gamma_{\mu_1}\nabla^{n - 1} \psi\right\}$ where $\nabla^{n - 1} := \nabla_{\mu_2}\nabla_{\mu_3}\cdots\nabla_{\mu_n}$ and the curly brackets denote the symmetric traceless part. Thus, the operator has irreducible or definite spin. In general twist two operators in a CFT dual to string theory on AdS are described in the large spin limit by rigidly rotating folded closed strings with large angular momentum \Gubser. Their anomalous dimensions (in the large spin limit) are given by calculating $E - J$ (in the large $J$ limit) for the spinning folded strings, where $E$ is energy and $J$ is angular momentum. See also \refs{\Alday,\ \KruczenskiA}.

A class of twist two operators in ${\cal N} = 4$ four dimensional $U(N)$ super Yang-Mills (SYM) was studied within the context of AdS/CFT correspondence in \Gubser. The operators are schematically of the form $\tr \left\{\phi\nabla_{\mu_1}\cdots \nabla_{\mu_k}\phi\right\}$. The theory is equivalent to string theory on $AdS_5\times S^5$. The 't Hooft coupling $\lambda = g^2N$ is related to the radius of curvature $R$ of $AdS_5$ in string units $l_s$ via the relation $\alpha'/R^2 \sim 1/\sqrt{\lambda}$ where $\alpha' = l_s^2$. At strong 't Hooft coupling $\lambda$ or at leading $\alpha'$ correction to the supergravity limit the anomalous dimensions, \ie, $\Delta - S - 2$, exhibit in the large spin limit similar logarithmic scaling behavior, \ie, $\Delta - S \approx \sqrt{\lambda}\ln S$. This and earlier observations led \Gubser\ to suggest on a general ground that in gauge theories the anomalous dimensions should have in the large $S$ limit the form $\Delta - S = f(\lambda)\ln S + {\cal O}(1/S)$ where $f(\lambda)$ is a function only of the 't Hooft coupling $\lambda$. The function $f(\lambda)$ interpolates between $\lambda$ at weak coupling and $\sqrt{\lambda}$ at strong coupling (up to numerical factors of order zero). In the study of Wilson loops with a light-like cusp the coefficient $f(\lambda)$ appears as a cusp anomalous dimension (up to a constant factor) \refs{\Korchemsky-\KorchemskyA, \ \Alday\KruczenskiA\Belitsky-\Basso}.

In the context of AdS/CFT correspondence, states in the CFT with higher twist are similarly described in AdS by rigidly rotating closed strings containing several light-like cusps or spiky foldings \Kruczenski.\foot{See \Burden\ for an earlier discussion on closed string configurations with cusps in relation to gravitational waves from heavy cosmic strings. See also \BurdenA\ for a discussion in relation to hadrons.} In the four dimensional $U(N)$ SYM, for example, the higher twist operators are schematically of the form ${\cal O}_m := \tr \left\{\nabla^{l_1}\phi_1\cdots \nabla^{l_m}\phi_m\right\}$ where the field $\phi_i$ can be either elementary or composite, and $\nabla^{l_m} := \nabla_{\mu_1}\nabla_{\mu_2}\cdots\nabla_{\mu_m}$. The total number of cusps or spikes of the spinning closed string is given by the twist of the corresponding operator or equivalently the total number of fields $\phi_i$ appearing in the operator. For example, the cusped closed string that describes the single trace operator ${\cal O}_m$ has $m$ cusps or spikes one for each constituent field $\phi_i$.  At strong t' Hooft coupling $\lambda$ and in the large spin $S$ limit their anomalous dimensions, \ie, $\Delta - S - m \approx \Delta - S$, are obtained in the context of the AdS/CFT correspondence by evaluating the quantity $E - J$ for the spinning cusped or spiky strings with $m$ cusps or spikes. The anomalous dimensions grow as $m\sqrt{\lambda}\ln S$ where $m$ is the twist or the number of fields that make up the operators or equivalently the number of spikes in the strings \Kruczenski. See also \refs{\Callan,\  \Belitsky}.

The spikes of the string foldings (and/or long strings that extend to or are near the boundary) are identified with soliton configurations of some integrable field theories \refs{\Jevicki,\ \JevickiK}. String theory on AdS spacetimes are related to integrable field theories such as the sinh-Gordon theory and its generalizations via (a construction analogous to) Pohlmeyer reduction \refs{\Pohlmeyer,\ \DeVega}. For example classical string theory on $AdS_2$ is equivalent to Liouville theory and classical string theory on $AdS_3$ is equivalent to the sinh-Gordon theory \refs{\LarsenA}. We later in the paper obtain the deformed sinh-Gordon theory that describes the spikes.

In this paper we study the structure of large angular momentum expansion of the quantity $E - J$ for certain spinning closed strings on ${\cal A}_3$. In the undeformed CFT it gives the anomalous dimensions of twist two and higher operators with large spin. We also compute the cusp anomalous dimension of a light-like Wilson loop. In a CFT or gauge theory the cusp anomalous dimension determines the coefficient $f(\lambda)$ in $E - J$ at large $J$. We also identify the analogous or deformed sinh-Gordon equation that describes the spikes or strings that reside near the boundary. This provides an example of single trace $T{\bar T}$ deformation in field theories which are not symmetric products. This will be useful to better understand the deformation in non-oribfold theories.

The paper is organized as follows. In section two we compute the quantity $E - J$ using the Nambu-Goto action for certain rotating folded and/or cusped closed strings. We demonstrate that it is bounded and has maximum finite value. The maximum value is determined in general by the coupling and the number of cusps. We also give remarkably simple Regge relations between $E$ and $J$ at a particular value of the deformation coupling. At this special value of the coupling, the string metric has, at fixed time coordinate, the topology of a cigar. In section three we discuss the cusp anomalous dimension of a light-like Wilson loop and its relation with the results obtained in section two. We find that in perturbation $E - J$ for a folded string agrees qualitatively (but not quantitatively) with the cusp anomalous dimension. Thus, we may still view it as a measure of anomalous dimension of certain operator with large spin (and twist two or higher) along the deformation in the coupling space. We comment on this later in the paper as we go along. In section four we determine the deformed sinh-Gordon model that describes the long strings residing near the boundary. In section five we formulate the main results and discuss future research directions.

\newsec{Rotating strings in ${\cal A}_3$}

We consider string theory on the background with the topology ${\cal A}_3\times S^3\times {\cal X}_4$ \AsratMHB. The background ${\cal A}_3$ interpolates between $AdS_3$ and a (an asymptotically) linear dilaton spacetime $\IR \times S^1\times \IR$. In the boundary it is equivalent to deforming the two dimensional conformal field theory (CFT) dual to (the long string sector of) string theory on $AdS_3\times S^3\times {\cal X}_4$ by an irrelevant operator of dimensions $(2, 2)$. The deformed worldsheet action on ${\cal A}_3$ in global coordinates is \AsratMHB,
\eqn\aaa{S_k = {k\over 2\pi}\int d^2 z \partial X^a\Sigma_{ab}{\bar\partial} X^b, \quad k\alpha'\Sigma_{ab} = G_{ab} + B_{ab},
}
where $X^a = (X^0, X^1, X^2) = (\tilde\varphi, \tilde\psi, \tilde\theta)$, $G_{ab}$ is the metric and $B_{ab}$ is the NSNS two form field. The angular coordinate $\tilde\psi$ has period $2\pi(1 - \gamma)$ \AsratMHB. For simplicity and convenience, we set the string coupling at $\gamma = 0$ to one, \ie, $g_s = 1$. The background fields $\Sigma_{ab}$, the dilaton $\phi$ and the NSNS three form flux $H$ are given by \AsratMHB
\eqn\bbb{\eqalign{
\Sigma_{22} & = 1,\cr
\Sigma_{11} & = e^{2\phi}\sinh^2\tilde\theta,\cr
\Sigma_{00} & = -e^{2\phi}\cosh^2\tilde\theta,\cr
\Sigma_{01} & = - \Sigma_{10} = -{1\over 2} e^{2\phi}\left[\gamma - \cosh(2\tilde\theta)\right],\cr
B & = B_{01}d\tilde\varphi\wedge d\tilde\psi, \quad B_{01} = k\alpha'\Sigma_{01},\cr
H & = dB = H_{012}d \tilde\varphi\wedge d\tilde\psi\wedge d\tilde\theta, \quad H_{012} = \partial B_{01}/\partial\tilde\theta,\cr
e^{2\phi} & = {1\over 1 + \gamma^2 - 2\gamma \cosh (2\tilde\theta)}.
}
}
The parameter $\gamma$ is dimensionless, and it is related to the dimensions $(2, 2)$ deformation coupling $\hat\gamma$ by $\hat\gamma = \gamma l_s^2$ (up to a constant factor). The coupling $\gamma$ takes \AsratMHB
\eqn\master{-1\leq \gamma \leq 0.
}
At $\gamma = 0$ the spacetime ${\cal A}_3$ is $AdS_3$ with radius $R = \sqrt{k\alpha'}$ and asymptotic string coupling $g_s = 1$. In the case $\gamma = -1$ the string background fields reduces to
\eqn\bbbbb{\eqalign{
ds^2 & = k\alpha'(d\tilde\theta^2 -{1\over 4}d\tilde\varphi^2 + {1\over 4}\tanh^2\tilde\theta d\tilde\psi^2),\cr
e^{2\phi} & = {1\over 4}{\rm sech}^2\tilde\theta,\cr
H &= 0.
}
}
As we will see shortly the quantities we study simplifies greatly in this special case. The metric has the topology of a product spacetime. At fixed time it has the topology of a cigar. Asymptotically it describes (with  $\tilde\psi$ uncompactified) a linear dilaton background $\IR^{(1, 1)}\times \IR$. In the Poincar\'e limit in which we take $-\gamma \ll 1$ and $\tilde\theta \gg 1$ with $\gamma e^{2\tilde\theta}$ fixed and finite we get 
\eqn\bbbbbbv{\eqalign{
ds^2 & = k\alpha'd\tilde\theta^2 + {k\alpha'\over 4}{e^{2\tilde\theta}\over 1 - \gamma e^{2\tilde\theta}}\left(-d\tilde\varphi^2 +  d\tilde\psi^2\right),\cr
e^{2\phi} & = {1\over 1 - \gamma e^{2\tilde\theta}},\cr
B & = {k\alpha'\over 4} {e^{2\tilde\theta}\over 1 - \gamma e^{2\tilde\theta}}d\tilde\varphi\wedge d\tilde\psi,\cr
H & = 2k\alpha' {e^{2\tilde\theta}\over (1 - \gamma e^{2\tilde\theta})^2}d \tilde\varphi\wedge d\tilde\psi\wedge d\tilde\theta,
}
}
Here the angular coordinate $\tilde\psi$ is uncompactified. For a detailed discussion of \aaa\ see \AsratMHB.

We now consider a closed string moving in the spacetime ${\cal A}_3$ \bbb\ with energy $E$ and angular momentum $J$. At $\gamma = 0$ or in $AdS_3$,  $E$ gives the scaling or mass dimension $\Delta$ and $J$ gives the spin $S$ of a certain single trace operator in the undeformed boundary CFT. Here the spacetime conformal symmetry is broken by the deformation and thus we cannot make such identifications. However, as we will show shortly, the combination $E - J$ agrees qualitatively (but not quantitatively) order by order in perturbation with the cusp anomalous dimension of a light-like Wilson loop. This suggests in general, in the large $J$ limit, we may still view $E - J$ as a measure of the anomalous dimension of certain (twist two or higher) operator with large spin along the flow or deformation in the coupling space. We leave a detailed discussion on this for a future work.

We describe the worldsheet embedding into the spacetime \bbb\ by
\eqn\bbbb{\tilde\varphi = \tau, \quad \tilde\psi = \omega\cdot \tau + (1 - \gamma) \cdot \kappa\cdot \sigma, \quad \tilde\theta = \tilde\theta(\sigma),
} 
where $\omega$ is the angular velocity of the string, and the variables $\sigma$ and $\tau$ parametrizes the worldsheet. $\sigma$ is the spatial coordinate along the string and $\tau$ describes its propagation in time. $\sigma$ is periodic with period $2\pi$. The integer $\kappa$ counts the number of times the string winds around the compact direction $\tilde\psi$. The winding number $\kappa$ can have either sign depending on the string orientation. For convenience we define the variable
\eqn\defin{\kappa_\gamma = (1 - \gamma)\kappa.
} 
The choice or parameterization \bbbb\ corresponds to a rigidly rotating closed string. In the case $\kappa = 0$ it corresponds to a folded rod like rotating closed string. In general we take the winding number $\kappa$ to be positive. The origin of the factor $1 - \gamma$ is to account or remind us for the absence of conical singularity in \bbb.

The dynamics of the classical string is described by the Nambu-Goto action. The Nambo-Goto action is given by
\eqn\ccc{S = -{\sqrt{\lambda}\over 2\pi}\int d\tau d\sigma {\cal L}, \quad R^2{\cal L} = \sqrt{-{\dot X}^2{X'}^2 + ({\dot X}\cdot X')^2},
}
where $\sqrt{\lambda} = R^2/\alpha'$ (in natural units) and ${\dot X}^2 = G_{ab}\partial_\tau X^a\partial_\tau X^b$, $X'^2 = G_{ab}\partial_\sigma X^a\partial_\sigma X^b$ and ${\dot X}\cdot X' = G_{ab}\partial_\tau X^a\partial_\sigma X^b$. For convenience we set $R = 1$ in string units. Thus, $\omega$ \bbbb\ is dimensionless. The string equations of motion are given by Euler's equations
\eqn\ddd{\partial_\mu {\partial {\cal L}\over\partial \partial_\mu X^a} - {\partial {\cal L}\over \partial X^a} = 0.
}
In terms of the metric components this gives
\eqn\dddd{\eqalign{
\partial_\sigma\left[{G_{ab}\left({X'}^b{\dot X}^2 - {\dot X}^b ({\dot X}\cdot X')\right)\over \sqrt{-{\dot X}^2{X'}^2 + ({\dot X}\cdot X')^2}}\right] + \partial_\tau\left[{G_{ab}\left({\dot X}^b{X'}^2 - {X'}^b ({\dot X}\cdot X')\right)\over \sqrt{-{\dot X}^2{X'}^2 + ({\dot X}\cdot X')^2}}\right]\cr
- {(\partial_a G_{bc})\left({\dot X}^b{\dot X}^c {X'}^2 + {X'}^b{X'}^c{\dot X}^2 - 2{\dot X}^b{X'}^c ({\dot X}\cdot {X'})\right)\over 2 \sqrt{-{\dot X}^2{X'}^2 + ({\dot X}\cdot X')^2}} = 0.
}
}

Using the metric \aaa\ and the parametrization \bbbb\ the Nambo-Goto Lagrangian density becomes
\eqn\cccc{{\cal L} = e^{\phi}\sqrt{{\tilde\theta}^{'2}(\cosh^2\tilde\theta - \omega^2 \sinh^2\tilde\theta)  + e^{2\phi} \kappa_\gamma^2\sinh^2\tilde\theta \cosh^2\tilde\theta}.
}

The equations of motion for $\tilde\varphi$ and $\tilde\psi$ give with $\kappa \neq 0$
\eqn\eee{{\tilde\theta'} = {\kappa_\gamma\over 2}\cdot {e^{2\phi}\sinh 2\tilde\theta\over e^{2\phi_0}\sinh 2\tilde\theta_0}\sqrt{e^{2\phi} \sinh^2 2\tilde\theta - e^{2(2\phi_0 -\phi)}\sinh^2 2\tilde\theta_0\over \cosh^2\tilde\theta - \omega^2 \sinh^2\tilde\theta},
}
where $\tilde\theta_0$ is an integration constant and $\phi_0 := \phi(\tilde\theta_0)$. The equation of motion for $\tilde\theta$ gives
\eqn\fff{\partial_\sigma\left[{\tilde\theta' {\dot X}^2\over {\cal L}}\right] - {\partial_{\tilde\theta} G_{\tilde\varphi\tilde\varphi} {X'}^2 + \partial_{\tilde\theta} G_{\tilde\psi\tilde\psi} \left[\omega^2 {X'}^2 + \kappa_\gamma^2{\dot X}^2 - 2\omega \kappa_\gamma ({\dot X}\cdot {X'})\right]\over 2{\cal L}} = 0,
}
where
\eqn\compon{\eqalign{
{\dot X}^2 & = -e^{2\phi}\left(\cosh^2\tilde\theta - \omega^2\sinh^2\tilde\theta\right),\cr
{X'}^2 & = \kappa_\gamma^2 \cdot e^{2\phi} \sinh^2\tilde\theta + {\tilde\theta}^{'2},\cr
{\dot X} \cdot {X'} & = \kappa_\gamma \cdot \omega e^{2\phi} \sinh^2\tilde\theta.
}
}
If we assume \eee, then \fff\ is satisfied.

We note that, in the case $\kappa \neq 0$, $\tilde\theta' = 0$ at $\tilde\theta = \tilde\theta_0$ and it diverges at $\tilde\theta = \tilde\theta_1$ where
\eqn\ggg{\tilde\theta_1 = \coth^{-1}\omega.
}
Note that $\tilde\theta_1$ is independent of the coupling $\gamma$. $\tilde\theta_0$ represents the turning point or valley and  $\tilde\theta_1$ represents the cusp or spike of a string segment, see Fig. 3. $\tilde\theta_0$ is the minimum radial distance. $\tilde\theta_1$ gives the radial position of the cusp or spike. In terms of $\tilde\theta_1 > \tilde\theta_0$ we rewrite \eee\ as
\eqn\gggg{{\tilde\theta'} = {\kappa_\gamma\over 2}\cdot \sinh\tilde\theta_1\cdot {e^{2\phi}\sinh 2\tilde\theta\over e^{2\phi_0}\sinh 2\tilde\theta_0}\sqrt{e^{2\phi} \sinh^2 2\tilde\theta - e^{2(2\phi_0 -\phi)}\sinh^2 2\tilde\theta_0\over \sinh(\tilde\theta_1 - \tilde\theta)\sinh(\tilde\theta_1 + \tilde\theta)}.
}
We rewrite the numerator in the square root as
\eqn\beforedoingGGG{\eqalign{
{e^{2\phi}\sinh^22\tilde\theta - e^{2(2\phi_0 - \phi)}\sinh^22\tilde\theta_0\over \sinh(\tilde\theta - \tilde\theta_0)\sinh(\tilde\theta + \tilde\theta_0)} & =\cr
{4\left[\left(1 + \gamma^2\right)\cosh (\tilde\theta + \tilde\theta_0) - 2\gamma \cosh(\tilde\theta - \tilde\theta_0)\right]\left[\left(1 + \gamma^2\right)\cosh(\tilde\theta - \tilde\theta_0) - 2\gamma\cosh(\tilde\theta + \tilde\theta_0)\right]\over \left(1 + \gamma^2 - 2\gamma \cosh2\tilde\theta\right)\left(1 + \gamma^2 - 2\gamma \cosh2\tilde\theta_0\right)^2}.
}
}
Note that near $\tilde\theta = \tilde\theta_0$, in general $\tilde\theta'$ goes to zero as $\sqrt{\tilde\theta - \tilde\theta_0}$. We also note that (for finite but large $\tilde\theta_1$ and nonzero $\gamma$) in general $\tilde\theta'$ diverges as $e^{\tilde\theta_1}/\sqrt{\tilde\theta_1 - \tilde\theta}$. For $\gamma = 0$ it instead diverges as $e^{4\tilde\theta_1}/\sqrt{\tilde\theta_1 - \tilde\theta}$.  In the case $\kappa = 0$, the radial coordinate $\tilde\theta$ extends from $\tilde\theta_0 = 0$ to a maximum value $\tilde\theta_1$ \ggg. For $\kappa = 0$ the closed string has four segments. In this paper we assume $\kappa$ is either one or zero. The choice $\kappa = 1$ is merely for simplicity and convenience.

The angle difference between the valley and spike is given by
\eqn\hhh{\Delta \theta = \int_{\tilde\theta_0}^{\tilde\theta_1} {d\tilde\theta\over \tilde\theta'} = {2\over \kappa_\gamma} \int_{\tilde\theta_0}^{\tilde\theta_1} {e^{2\phi_0}\sinh 2\tilde\theta_0\over e^{2\phi} \sinh 2\tilde\theta}\sqrt{\cosh^2\tilde\theta - \omega^2 \sinh^2\tilde\theta \over e^{2\phi} \sinh^2 2\tilde\theta - e^{2(2\phi_0-\phi)}\sinh^2 2\tilde\theta_0}d\tilde\theta.
}
The cusps or spikes are equally spaced and the components of the metric \bbb\ are independent of the angular coordinate $\tilde\psi$. Thus, $\Delta\theta$ satisfy
\eqn\constrainTH{\Delta\theta = {1\over 2}\cdot{2\pi\over n},
}
where $n \geq 2$ is the number of cusps/spikes or twist. This fixes $\tilde\theta_0 = \tilde\theta_0(n, \omega, \kappa, \gamma)$.

The conserved densities are given by
\eqn\hhhh{J_a^\tau = {\partial {\cal L}\over \partial \partial_\tau X^a} = {G_{ab}\left[-{\dot X}^b{X'}^2 + {X'}^b\left({\dot X}\cdot {X'}\right)\right]\over {\cal L}}.
}
The angular momentum carried by the string is then given by
\eqn\iii{\eqalign{
J & = \kappa\cdot 2n\cdot \int_{\tilde\theta_0}^{\tilde\theta_1} {d\tilde\theta\over \tilde\theta'}\left( - {\sqrt{\lambda}\over 2\pi} \cdot J_{\tilde\psi}^\tau\right), \cr
& =  \kappa\cdot 2n\cdot \int_{\tilde\theta_0}^{\tilde\theta_1} {d\tilde\theta\over \tilde\theta'}\left( - {\sqrt{\lambda}\over 2\pi} \cdot {-\omega e^{2\phi}\sinh^2\tilde\theta \over {\cal L}}{\tilde\theta}^{'2}\right),\cr
& = \kappa\cdot 2n\cdot  {\sqrt{\lambda}\over 2\pi} \int_{\tilde\theta_0}^{\tilde\theta_1} {\omega \sinh^2\tilde\theta\over \sinh 2\tilde\theta}\sqrt{e^{2\phi}\sinh^2 2\tilde\theta - e^{2(2\phi_0 -\phi)}\sinh^2 2\tilde\theta_0\over \cosh^2\tilde\theta - \omega^2 \sinh^2 \tilde\theta}d\tilde\theta.
}
}
Each string segment contributes equally and thus the factor $2n$ in front of the integral. In the case $\kappa = 0$ we have
\eqn\iiii{J = 4\cdot  {\sqrt{\lambda}\over 2\pi} \int_{\tilde\theta_0}^{\tilde\theta_1} {e^{\phi}\omega \sinh^2\tilde\theta\over\sqrt{\cosh^2\tilde\theta - \omega^2 \sinh^2 \tilde\theta}}d\tilde\theta.
}
The factor $4$ in front of the integral counts the four segments. We rewrite the integral \iiii\ as
\eqn\iiiii{J = {\omega\sqrt{\lambda}\over 2\pi}\sqrt{x_1\over -\gamma}\int_{x_0}^{x_1} dx \sqrt{x\over (x_1 - x)(x + 1)(x + \alpha)},
}
where
\eqn\iiiiipara{x_0 = 0, \quad x_1 = \sinh^2\tilde\theta_1 = {1\over \omega^2 - 1}, \quad \alpha = {(1 - \gamma)^2\over -4\gamma}.
}
The integral \iiiii\ is expressible in terms of elliptic functions. The exact result is given by
\eqn\iiiiii{J = {2\sqrt{\lambda}\over\pi(1 - \gamma)}\left\{\Pi\left[{1\over \omega^2}, \left({1 + \gamma\over 1 - \gamma}\right)^2\cdot {1\over \omega^2}\right] - K\left[\left({1 + \gamma\over 1 - \gamma}\right)^2\cdot {1\over \omega^2}\right]\right\}.
}
where $\Pi(\alpha, \beta)$ is the complete elliptic integral of the third kind and $K(\alpha)$ is the complete elliptic integral of the first kind, see appendix A for our convention. In the special case where $\gamma = -1$ the expression \iiiiii\ simplifies greatly. In fact we get the simple expression
\eqn\iiiiiii{J = {\sqrt{\lambda}\over 2}\left({\omega\over \sqrt{\omega^2 - 1}} - 1\right).
}
The general and exact solution to the angular momentum \iii\ will be given momentarily. The spacetime energy carried by the string is similarly given by
\eqn\jjj{\eqalign{
E & =  \kappa\cdot 2n\cdot \int_{\tilde\theta_0}^{\tilde\theta_1} {d\tilde\theta\over \tilde\theta'}\left( - {\sqrt{\lambda}\over 2\pi} \cdot J_t^\tau \right),\cr
& =  \kappa\cdot 2n\cdot \int_{\tilde\theta_0}^{\tilde\theta_1} {d\tilde\theta\over \tilde\theta'}\left[ {\sqrt{\lambda}\over 2\pi} \cdot {e^{2\phi}\cosh^2\tilde\theta\left(e^{2\phi}\kappa_\gamma^2\sinh^2\tilde\theta + {\tilde\theta}^{'2}\right)\over {\cal L}}\right],\cr
& =  \kappa \cdot 2n\cdot {\sqrt{\lambda}\over 2\pi} \int_{\tilde\theta_0}^{\tilde\theta_1} {(e^{2\phi}\kappa_\gamma^2\sinh^2\tilde\theta + {\tilde\theta}^{'2})e^{2\phi_0}\sinh 2\tilde\theta_0\over 2 \kappa_\gamma e^{2\phi}\sinh^2\tilde\theta}{d\tilde\theta\over \tilde\theta'}.
}
}
In the case $\kappa = 0$ we have
\eqn\jjjj{E  = 4\cdot  {\sqrt{\lambda}\over 2\pi} \int_{\tilde\theta_0}^{\tilde\theta_1} {e^{\phi} \cosh^2\tilde\theta\over\sqrt{\cosh^2\tilde\theta - \omega^2 \sinh^2 \tilde\theta}}d\tilde\theta.
}
The factor $4$ in front of the integral is the total number of segments of the folded rodlike string. We rewrite the expression \jjjj\ as
\eqn\jjjjj{E = {\sqrt{\lambda}\over 2\pi}\sqrt{x_1\over -\gamma}\int_{x_0}^{x_1}dx\sqrt{x + 1\over x(x_1 - x)(x + \alpha)},
}
where the variables $x_0, x_1$ and $\alpha$ are given above in \iiiiipara. The result to the integral is
\eqn\jjjjjj{E = {2\sqrt{\lambda}\over\pi(1 - \gamma)\omega}\Pi\left[{1\over \omega^2}, \left({1 + \gamma\over 1 - \gamma}\right)^2\cdot {1\over \omega^2}\right].
}
In the case $\gamma = -1$ it reduces to
\eqn\jjjjjjj{E = {\sqrt{\lambda}\over 2\sqrt{\omega^2 - 1}}.
}
We will give the general and exact solution to \jjj\ in terms of the angular momentum $J$ shortly when we discuss $E - J$ or equivalently at $\gamma = 0$ the anomalous dimensions of certain higher twist operators. 

The angular momentum and energy are related as
\eqn\kkk{\eqalign{
E - \omega J & = \kappa\cdot 2n \cdot \int_{\tilde\theta_0}^{\tilde\theta_1} {d\tilde\theta\over {\tilde\theta'}}\left({\sqrt{\lambda}\over 2\pi}\cdot {\cal L}\right),\cr
&  = \kappa\cdot 2n\cdot {\sqrt{\lambda}\over 2\pi}\int_{\tilde\theta_0}^{\tilde\theta_1}e^{2\phi}\sinh 2\tilde\theta \sqrt{\cosh^2\tilde\theta - \omega^2 \sinh^2\tilde\theta\over e^{2\phi}\sinh^2 2\tilde\theta - e^{2(2\phi_0 -\phi)}\sinh^2 2\tilde\theta_0}d\tilde\theta.
}
}
In the case $\kappa = 0$ we instead have
\eqn\kkkk{E - \omega J = 4\cdot {\sqrt{\lambda}\over 2\pi}\int_{\tilde\theta_0}^{\tilde\theta_1}e^{\phi} \sqrt{\cosh^2\tilde\theta - \omega^2 \sinh^2\tilde\theta}d\tilde\theta.
}
We rewrite \kkkk\ as
\eqn\kkkkk{E - \omega J = {1\over 2\pi}\sqrt{\lambda\over -\gamma x_1}\int_{x_0}^{x_1}dx\sqrt{x_1 - x\over x(x+1)(x + \alpha)}, 
}
where the variables $x_0, x_1$ and $\alpha$ are defined above in \iiiiipara. We get
\eqn\kkkkkk{\eqalign{
E - \omega J & = {\alpha + x_1\over \pi}\sqrt{\lambda\over -\gamma\alpha x_1(x_1 + 1)}\left[K\left(x_1(\alpha - 1)\over \alpha(x_1 + 1)\right) - \Pi\left(-{x_1\over \alpha}, {x_1(\alpha - 1)\over \alpha(x_1 + 1)}\right)\right],\cr
& = {\sqrt{\lambda}\over 2\pi}{(1 + \gamma)^2 - \omega^2(1 - \gamma)^2\over \omega \gamma (1 - \gamma)}\cr
& \cdot \left\{K\left[\left({1 + \gamma\over 1 - \gamma}\right)^2\cdot {1\over \omega^2}\right] - \Pi\left[{4\gamma\over (\omega^2 - 1)(1 - \gamma)^2}, \left({1 + \gamma\over 1 - \gamma}\right)^2\cdot {1\over \omega^2}\right]\right\}.
}
}
In the special case $\gamma = -1$ the energy-momentum relation \kkkkkk\ simplifies to 
\eqn\kkkkkkkkk{E - \omega J = {\sqrt{\lambda}\omega\over 2}\left(1 - {\sqrt{\omega^2 - 1}\over \omega}\right).
}
Using either the result \iiiiiii\ or \jjjjjjj\ we get the following exact and simple relation between the energy and angular momentum in the case $\gamma = -1$. A straightforward simplification gives
\eqn\kkkkkkkkkk{E = \left(\omega  + \sqrt{\omega^2 - 1}\right)J.
}
We will discuss the $\omega$ dependence of $J$ in a moment in the next subsection. The equation \kkkkkkkkkk\ implies the Regge relation 
\eqn\regge{m^2 = l(l + 2),
}
where the dimensionless energy $m$ and angular momentum $l$ are defined as 
\eqn\renormal{
m:= {2E\over \sqrt{\lambda}}, \quad l:= {2J\over \sqrt{\lambda}}.
}
We note that for small $l \ll 2$ the Regge relation \regge\ is that of flat spacetime with a different slope. In flat spacetime we have $m^2 = 4l$. However, here we have $m^2 = 2l$. The exact solution to the integral \kkk\ will be given shortly. 

In what follows we begin our discussion with folded rodlike rotating strings, \ie, $n = 2$.

\subsec{Folded rodlike rotating strings}

In the context of AdS/CFT folded rigidly rotating closed strings describe the anomalous dimensions of twist two operators \Gubser. Therefore, at $\gamma = 0$, the quantity $E - J$ gives the anomalous dimensions of certain twist two operators in the undeformed CFT.  

The angular momentum $J$ is given by \iiiiii\ which we write agian below for convenience. It is given by
\eqn\twaaa{J = {2\sqrt{\lambda}\over\pi(1 - \gamma)}\left\{\Pi\left[{1\over \omega^2}, \left({1 + \gamma\over 1 - \gamma}\right)^2\cdot {1\over \omega^2}\right] - K\left[\left({1 + \gamma\over 1 - \gamma}\right)^2\cdot {1\over \omega^2}\right]\right\}.
}
The energy $E$ is given by \jjjjjj\ which we also write agian below. It is given by
\eqn\twbbb{E = {2\sqrt{\lambda}\over\pi(1 - \gamma)\omega}\Pi\left[{1\over \omega^2}, \left({1 + \gamma\over 1 - \gamma}\right)^2\cdot {1\over \omega^2}\right].
}
The expression for the combination $E - \omega J$ which we discuss shortly is given earlier in \kkkkkk. See Fig. 1 for plots of the energy $E$ as a function of the angular momentum $J$ for various values of the deformation coupling $\gamma$. We note that for large values of $J$ and in the cases in which $\gamma \neq 0$ the energy $E$ increases linearly. Thus, in general we expect a relation of the form $E = J + \sqrt{\lambda}\cdot \delta(\gamma)$. We will determine the asymptotic intercept $\delta(\gamma)$ in a moment. In the case $\gamma = -1$ we have the simple exact result \regge. In the figure it is represented by the red curve.

\ifig\loc{The plots depict the energy $E$ of the rigidly rotating folded closed string as a function of the angular momentum $J$ in units of $\sqrt{\lambda}$ for several values of the coupling $\gamma$. The brown curve is for $\gamma = 0$, the orange curve is for $\gamma = -1/4$, the purple curve is for $\gamma = -1/2$ and the red curve is for $\gamma = -1$. The dashed gray curve represents $E = J$.}
{\epsfxsize4.2in\epsfbox{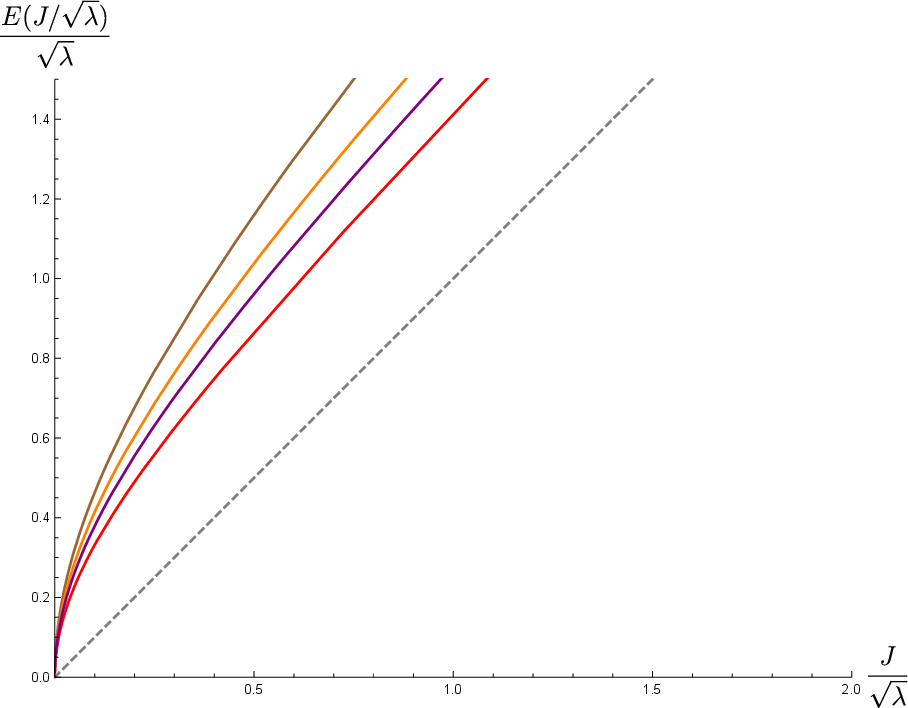}}

We are in particular interested in the case $\omega$ is close to one. This is equivalent to a folded rodlike string that extends to the boundary, see \ggg. Thus, we write $\omega = 1 + 2\eta$ where $0 < \eta \ll 1$.

In the case $\gamma = 0$ we find in perturbation theory
\eqn\twccc{J = {\sqrt{\lambda}\over2\pi} \left[{1\over \eta} + 2 + \ln\left({\eta\over 4}\right)\right] + {\cal O}(\eta),
}
\eqn\twddd{E = {\sqrt{\lambda}\over2\pi} \left[{1\over \eta}  - \ln\left({\eta\over 4}\right)\right] + {\cal O}(\eta),
}
and 
\eqn\twemjdd{E - (1 + 2\eta)J =  {\sqrt{\lambda}\over \pi}\ln\left({1\over \eta}\right) + {\cal O}(\eta). 
}
Therefore, the anomalous dimension of the corresponding twist two operator in the undeformed CFT is given in the large spin limit, \ie, $S\gg\sqrt{\lambda}$, by
\eqn\tweee{\Delta - S = {\sqrt{\lambda}\over \pi}\ln\left({S\over \sqrt{\lambda}}\right) + {\cal O}({\sqrt{\lambda}/S}).
}
This is the result obtained earlier in \Gubser.

We next consider $\gamma \neq 0$ and $0 < -\gamma \ll 1$. We write $\omega$ as $\omega = 1 + 2\eta$. We take $0 < \eta \ll 1$ and $J \gg \sqrt{\lambda}$. We first do perturbation in $\gamma$. Following this, we perform perturbation in $\eta$. Thus, we effectively assume $0 < -\gamma/\eta \ll 1$. The angular momentum $J$ in perturbation theory is given by
\eqn\twjfff{\eqalign{
{\eta\cdot J\over \sqrt{\lambda}} & = {1\over 2\pi} + {\eta\ln(\eta)\over 2\pi } + {\eta^2 \ln(\eta)\over 4\pi } - {3\eta^3\ln(\eta)\over 8\pi } + {\cal O}\left(\eta^4\right)\cr
& + \left({\gamma\over \eta}\right)\left[{1\over 6\pi} - {(13 + 6 \ln(\eta))\over 24\pi}\eta^2 + {\eta^3\over 2\pi} + {(-47 + 12\ln(\eta))\over 128\pi}\eta^4 + {\cal O}\left(\eta^6\right)\right]\cr
& + \left({\gamma\over \eta}\right)^2\left[{1\over 10\pi } + {3\eta\over 20\pi} + {\eta^2\over 8\pi} + {\eta^3\ln(\eta)\over 8\pi } - {\eta^4\ln(\eta)\over 32\pi } + {3\eta^5\ln(\eta)\over 64\pi } + {\cal O}\left(\eta^6\right)\right] + {\cal O}(\gamma^3).
}
}
We are keeping terms only relevant for our following discussion. Thus, up to and including order two in $\gamma$ we have
\eqn\twjfff{
J = {\sqrt{\lambda}\over 2\pi}\left({1\over \eta} + \ln \eta\right) + {\sqrt{\lambda}\over 2\pi}\left({1\over 3 \eta^2} - {1\over 2}\ln\eta\right)\gamma + {\sqrt{\lambda}\over 2\pi}\left({1\over 5\eta^3} + {3\over 10\eta^2} +{1\over 4\eta}+{1\over4}\ln\eta\right)\gamma^2 + {\cal O}(\gamma^3).
}
The energy carried by the spinning string is given in perturbation by
\eqn\twefff{\eqalign{
E & = {\sqrt{\lambda}\over 2\pi}\left({1\over \eta} - \ln \eta + {\eta \ln \eta \over 2} - {\eta^2 \ln \eta \over 4} + {\eta^3 \ln \eta \over 16} + {\cal O}(\eta^4) \right) \cr
& +  {\sqrt{\lambda}\over 2\pi}\left({1\over 3 \eta} + {4\over 3} + {1\over 4}\eta - {1\over 2}\eta\ln\eta + {(17 + 12 \ln\eta)\over 64}\eta^3 -{1\over 4}\eta^4\ln\eta+ {\cal O}(\eta^5)\right)\left({\gamma\over \eta}\right) \cr
& + {\sqrt{\lambda}\over 2\pi}\left({1\over 5 \eta} + {9\over 10} + {29\eta\over 20} - {1\over 4}\eta^2\ln\eta - {1\over 16}\eta^3\ln\eta + {1\over 32}\eta^4\ln\eta + {\cal O}(\eta^5)\right)\left({\gamma\over \eta}\right)^2 + {\cal O}(\gamma)^3
}
}
Thus, up to and including order two in $\gamma$ we have
\eqn\tweXefff{
E = {\sqrt{\lambda}\over 2\pi}\left({1\over \eta} - \ln \eta\right) + {\sqrt{\lambda}\over 2\pi}\left({1\over 3 \eta^2} + {4\over 3\eta} - {1\over 2}\ln\eta\right)\gamma + {\sqrt{\lambda}\over 2\pi}\left({1\over 5\eta^3} + {9\over 10 \eta^2} + {29\over 20 \eta} - {1\over 4}\ln\eta\right)\gamma^2 + {\cal O}(\gamma^3).
}
Similarly, the combination $E - \omega J$ given in \kkkkkk\ becomes in perturbation
\eqn\twemjfff{\eqalign{
{E - (1 + 2\eta)J\over \sqrt{\lambda}} & = -{1\over \pi}\ln(\eta) - {\eta\over \pi}\ln(\eta) - {\eta^2\over 4\pi}\ln(\eta) + {\cal O}(\eta^3)\cr
& + \left({\gamma\over \eta}\right)\left[{1\over 3\pi } + {2\eta\over 3\pi} +  {\eta^2\ln(\eta)\over 2\pi} - {\eta^3\over 2\pi} + {\cal O}(\eta^4)\right]\cr
& + \left({\gamma\over \eta}\right)^2\left[{1\over 10\pi } + {3\eta\over 10\pi } - {\eta^2\ln(\eta)\over 4\pi} - {\eta^3\ln(\eta)\over 4\pi} + {\eta^4\ln(\eta)\over 32\pi}+ {\cal O}(\eta^5)\right] + {\cal O}(\gamma^3).
}
}
This is consistent with \twjfff\ and \twefff. Thus, up to and including order two in $\gamma$ we have
\eqn\twemjfffx{\eqalign{
E - J & = {\sqrt{\lambda}\over \pi}\ln\left({1\over \eta}\right) + {2\gamma\sqrt{\lambda}\over 3\pi\eta} + {\sqrt{\lambda}\over 2\pi}\left({3\over 5\eta^2} + {6\over 5\eta} - {1\over 2}\ln\eta\right)\gamma^2 + {\cal O}(\gamma^3).\cr
E - (1 + 2\eta) J &= {\sqrt{\lambda}\over \pi}\ln\left({1\over \eta}\right) + {\gamma\sqrt{\lambda}\over 3\pi\eta} + {\sqrt{\lambda}\over 2\pi}\left({1\over 5\eta^2} + {3\over 5\eta} - {\ln\eta\over 2}\right)\gamma^2 + {\cal O}(\gamma^3).
}
}
Therefore, since $\gamma < 0$ the first correction is negative.

In the large angular momentum limit, \ie, $J\gg \sqrt{\lambda}$, we therefore get using \twjfff\ and the first equation in \twemjfffx 
\eqn\twfff{\eqalign{
{1\over \eta}& = 2\pi {J\over \sqrt{\lambda}} \left[1 - {2\pi\over 3}{J\over \sqrt{\lambda}}\gamma + {4\pi^2\over 45}{J^2\over \lambda}\gamma^2 + {\cal O}\left({J^3\over \lambda^{3/2}}\gamma^3\right)\right],  \cr
E - J & = {\sqrt{\lambda}\over \pi}\left[\ln\left(2\pi {J\over \sqrt{\lambda}}\right) + {2\pi \over 3}{J\over \sqrt{\lambda}}\gamma + {8\pi^2\over 45}{J^2\over \lambda}\gamma^2  + {\cal O}\left({J^3\over \lambda^{3/2}}\gamma^3\right)\right].
}
}
We note that the first non vanishing correction comes at order $\gamma$ and it is negative. We also note that the radial distance of the tip of the folded string is 
\eqn\len{\tilde\theta_1 =  {1\over 2} \left[\ln\left(2\pi {J\over \sqrt{\lambda}}\right) - {2\pi \over 3}{J\over \sqrt{\lambda}}\gamma - {2\pi^2\over 15}{J^2\over \lambda}\gamma^2  + {\cal O}\left({J^3\over \lambda^{3/2}}\gamma^3\right)\right].
}
Thus, the deformation for $\gamma < 0$ stretches the strings or equivalently pushes away the conformal boundary \AsratMHB.

In the case $\gamma$ is finite and $\omega = 1 + 2\eta \approx 1$ \ie, $J \gg \sqrt{\lambda}$, \kkkkkk\ gives
\eqn\twtomegaone{E - J = {2\sqrt{\lambda}\over \pi(1- \gamma)}K\left(\left({1 + \gamma\over 1 - \gamma}\right)^2\right).
}
This is consistent with \twaaa\ and \twbbb. This gives the asymptotic intercept $\sqrt{\lambda}\cdot \delta(\gamma)$. In the small $\gamma$ limit it reduces to 
\eqn\twtomegaonex{{E - J\over \sqrt{\lambda}} = {1\over \pi}\ln\left({-1\over \gamma}\right) + {\cal O}\left({\gamma^2}\right).
}
At $\gamma = -1$ we have
\eqn\upperbminussx{{E - J\over \sqrt{\lambda}} = {1\over 2},
}
which is in agreement with the result \kkkkkkkkk, or \regge. Therefore, for $\gamma < 0$ the quantities $E - J$ are bounded from above. Also note that the bound or the maximum value is the asymptotic intercept $\sqrt{\lambda}\cdot \delta(\gamma)$ in Fig. 1. In general, in the large angular momentum limit, we have \twtomegaone
\eqn\lfilaton{E \geq J + {\sqrt{\lambda}\over 2}.
}
See Fig. 2 for plots of $E - \omega J$ \kkkkkk\ for different values of the deformation coupling $\gamma$.

\ifig\loc{The plots depict the quantity $E - \omega J$ as a function of the angular momentum $J$ in units of $\sqrt{\lambda}$ for different values of the coupling $\gamma$. The orange curve is for $\gamma = -1/4$, the purple curve is for $\gamma = -1/2$ and the red curve is for $\gamma = -1$. The gray dashed lines represent the maximum values of $E - \omega J$ or the asymptotic intercepts $\sqrt{\lambda}\cdot \delta(\gamma)$ per $\sqrt{\lambda}$ \twtomegaone.}
{\epsfxsize4.5in\epsfbox{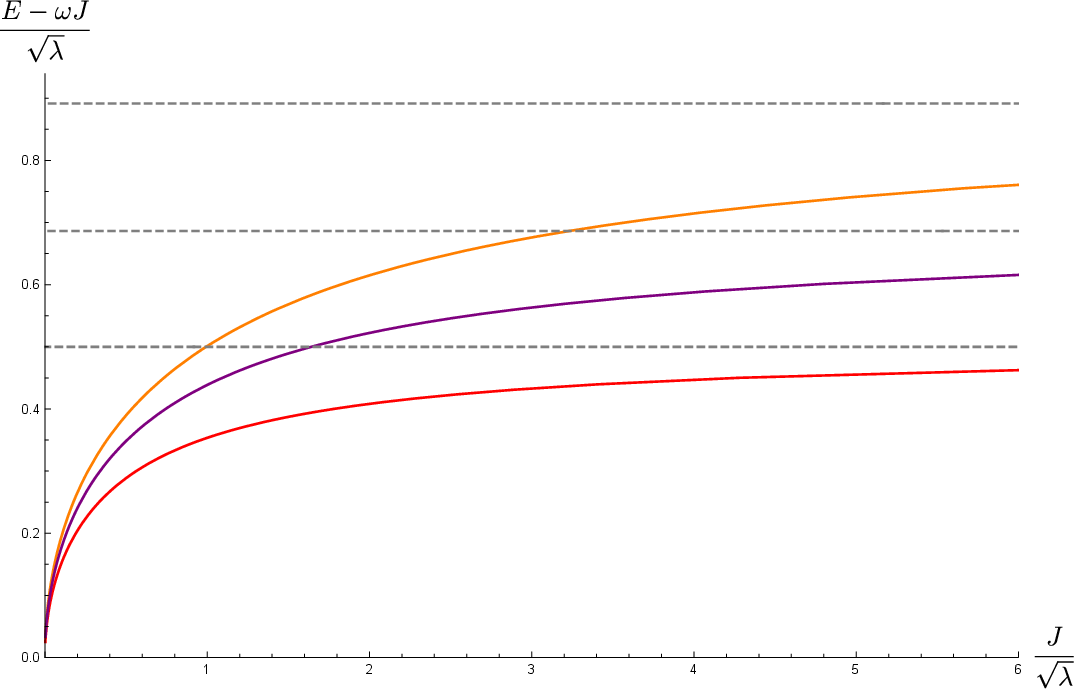}}

To study short strings we need to consider the case where $\omega$ is large, \ie, $\omega \gg 1$. We find
\eqn\twtlomega{\eqalign{
J & = {\sqrt{\lambda}\over 2(1 - \gamma)}{1\over \omega^2} + {\cal O}\left({1\over \omega^4}\right),\cr
E& = {\sqrt{\lambda}\over 1 - \gamma}{1\over \omega} + {\cal O}\left({1\over \omega^3}\right).
}
}
Thus, the energy is given by
\eqn\twtlomega{
E = \sqrt{{2\over 1 - \gamma}}\lambda^{1\over 4}J^{1\over 2} + {\cal O}(\lambda^{3/4}J^{3/2}).
}
Note in the short string case the power of $\lambda$ is $1/4$. We also note that at $\gamma = 0$ \twtlomega\ gives
\eqn\twtlomegaaaa{
\Delta^2 = 2\lambda^{1\over 2}S,
}
which is the result in flat spacetime.

In the case $\gamma = -1$ we have the exact result \kkkkkkkkkk,
\eqn\twtlomegagminus{E^2 = J(J + \lambda^{1\over 2}).
}
We note that the leading exponent of $J$ at large values is two. In flat spacetimes however we have one \twtlomegaaaa. 

\subsec{Cusped rotating strings}

We now extend the discussion of the previous subsection on the quantity $E - J$ to rigidly rotating cusped closed strings. In the context of AdS/CFT, spinning cusped strings describe higher twist operators \Kruczenski. That is, at $\gamma = 0$, $E - J$ gives the anomalous dimensions of certain higher twist operators.

For convenience and simplicity we set the winding $\kappa = 1$. In the former discussion $\tilde\theta_0 = 0$. Here it is fixed by the equation \hhh. We have from $\hhh$ with $\kappa = 1$
\eqn\twhaa{\Delta \theta = \int_{\tilde\theta_0}^{\tilde\theta_1} {d\tilde\theta\over \tilde\theta'} = {2\over 1 - \gamma} \int_{\tilde\theta_0}^{\tilde\theta_1} {e^{2\phi_0}\sinh 2\tilde\theta_0\over e^{2\phi} \sinh 2\tilde\theta}\sqrt{\cosh^2\tilde\theta - \omega^2 \sinh^2\tilde\theta \over e^{2\phi} \sinh^2 2\tilde\theta - e^{2(2\phi_0-\phi)}\sinh^2 2\tilde\theta_0}d\tilde\theta = {1\over 2}\cdot{2\pi\over n}.
}
At $\gamma = 0$ $n$ is the twist. We recall that $\Delta\theta$ is the angle difference between the cusp or spike and the valley or the closest point to the coordinates origin. We rewrite the integral as
\eqn\integralexact{\eqalign{
\Delta \theta & ={1\over 1 - \gamma} \sqrt{(x_0^2 - 1)(1 + \gamma^2)\over (x_1 - 1)\zeta_0[\zeta_0(2x_0 + \zeta_0) + 1]}\int_{x_0}^{x_1}dx \left({x+\zeta_0\over x^2 - 1}\right)\sqrt{(x_1 - x)(x + \zeta_0)\over (x - x_0)(x + \zeta_1)}.
}
}
where
\eqn\cinstsevjjtheta{x_0 = \cosh 2\tilde\theta_0, \quad x_1 = \cosh2\tilde\theta_1 = {\omega^2 + 1\over \omega^2 - 1}, \quad \zeta_0 = {1 + \gamma^2\over -2\gamma}, \quad \zeta_1 = {\zeta_0(2 + x_0\zeta_0) + x_0\over \zeta_0(\zeta_0 + 2x_0) + 1}.  
}
Note that $\zeta_0 \geq \zeta_1$. In general we find
\eqn\integralthetaiiii{\eqalign{
\Delta \theta & = {1\over 1 - \gamma}\sqrt{(x_0^2 - 1)(1 + \gamma^2)\over (x_1 - 1)\zeta_0[\zeta_0(2x_0 + \zeta_0) + 1]}\cr
& \cdot \frac{-\zeta _0-x_1}{\sqrt{\left(\zeta _0+x_0\right) \left(\zeta _1+x_1\right)}}\left[2 \Pi \left(\frac{x_0-x_1}{x_0+\zeta _0}, \frac{\left(x_0-x_1\right) \left(\zeta _1-\zeta _0\right)}{\left(x_0+\zeta _0\right) \left(x_1+\zeta _1\right)}\right)+ \right.\cr
& \left(\zeta _0-1\right) \Pi \left(-\frac{\left(x_0-x_1\right) \left(\zeta _0-1\right)}{\left(x_1+1\right) \left(x_0+\zeta _0\right)}, \frac{\left(x_0-x_1\right) \left(\zeta _1-\zeta _0\right)}{\left(x_0+\zeta _0\right) \left(x_1+\zeta _1\right)}\right)- \cr
& \left.\left(\zeta _0+1\right) \Pi \left(-\frac{\left(x_0-x_1\right) \left(\zeta _0+1\right)}{\left(x_1-1\right) \left(x_0+\zeta _0\right)}, \frac{\left(x_0-x_1\right) \left(\zeta _1-\zeta _0\right)}{\left(x_0+\zeta _0\right) \left(x_1+\zeta _1\right)}\right)\right] = {\pi \over n}, \quad n > 2,
}
}
where $\Pi(\alpha, \beta)$ is the complete elliptic integral of the third kind, see appendix A. $\Delta\theta$ determines $x_0 = \cosh2\tilde\theta_0$ in terms of the variables $n, \omega, \gamma$. See Fig. 3 for an example of a closed string configuration with five spikes. 
\ifig\loc{A closed string with five cusps or spikes. We chose $\gamma = -1$ and $\omega = 1.001$. We have normalized the radial distance $\tilde\theta_1$ to one.}
{\epsfxsize4.0in\epsfbox{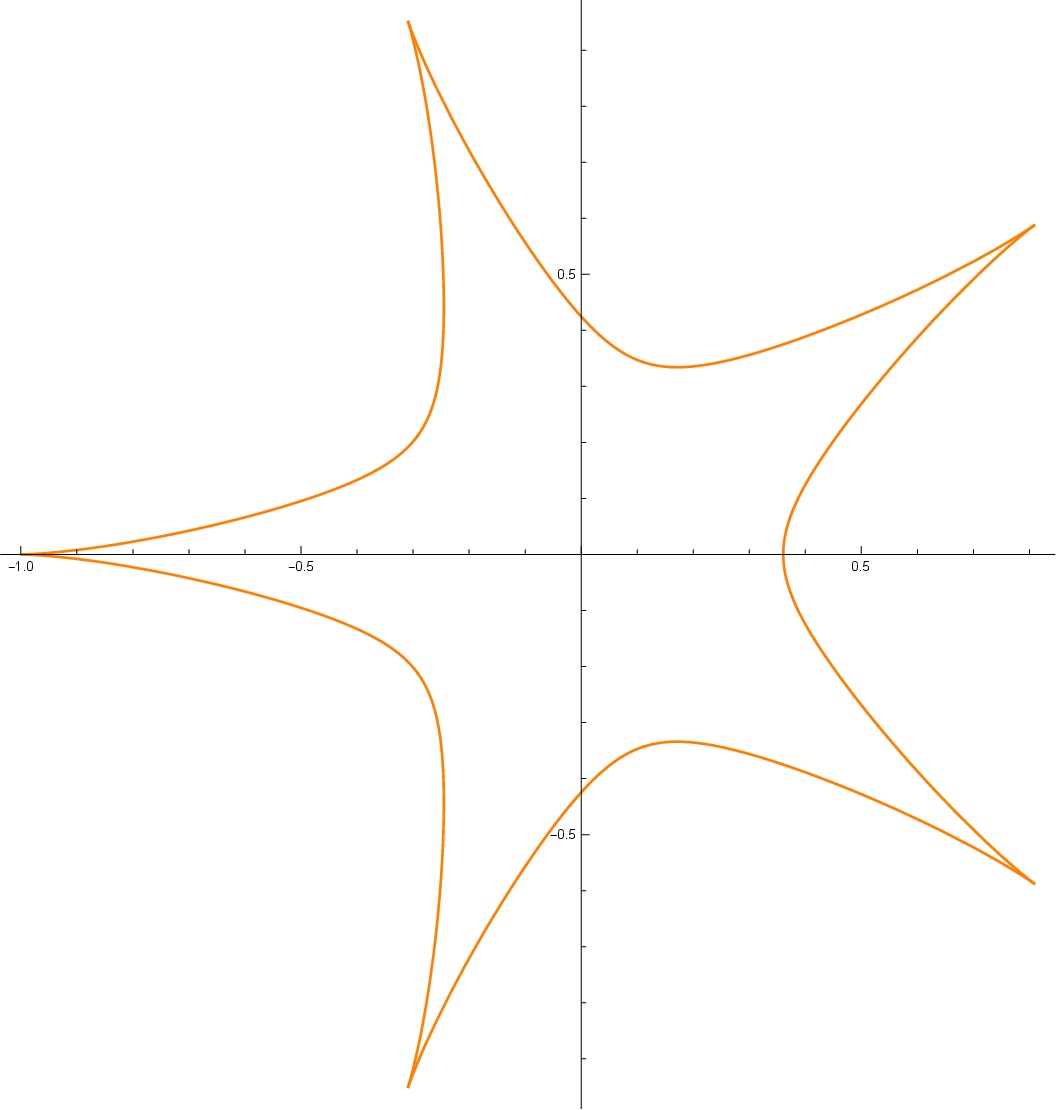}}

\ifig\loc{The plots depict the quantity $E - \omega J$ as a function of the angular momentum $J$ in units of $\sqrt{\lambda}$ in the case $\gamma = -1$ for different number $n$ of cusps or spikes. The brown curve is for $n = 3$, the red curve is for $n = 4$, the magenta curve is for $n = 8$ and the orange curve is for $n = 1000$. The horizontal dashed gray line represents the maximum value (2.107) of $E - \omega J$. The vertical dashed gray lines represent the maximum values (2.94) of $J$.}
{\epsfxsize5.2in\epsfbox{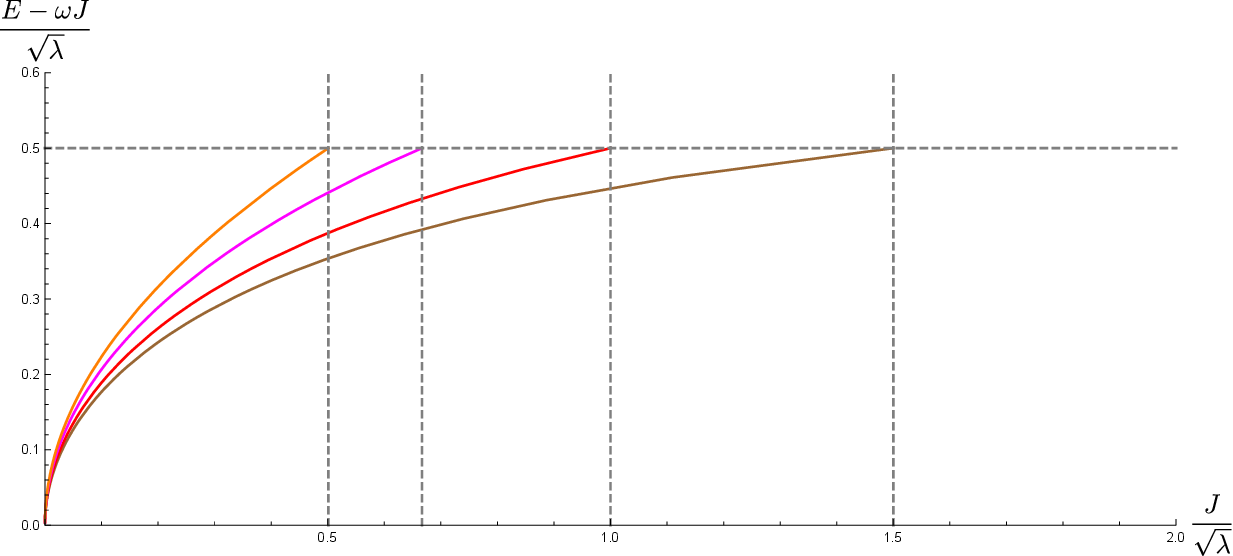}}

\ifig\loc{The plots depict the quantity $E - \omega J$ as a function of the angular momentum $J$ in units of $\sqrt{\lambda}$ in the case $n = 5$ for different values of the coupling. The orange curve is for $\gamma = -1/2$, the brown curve is for $\gamma = -3/4$ and the red curve is for $\gamma = -1$. The gray dashed horizontal lines represent the maximum values (2.107) of $E - \omega J$. The gray dashed vertical lines represent the maximum values (2.94) of $J$.}
{\epsfxsize4.8in\epsfbox{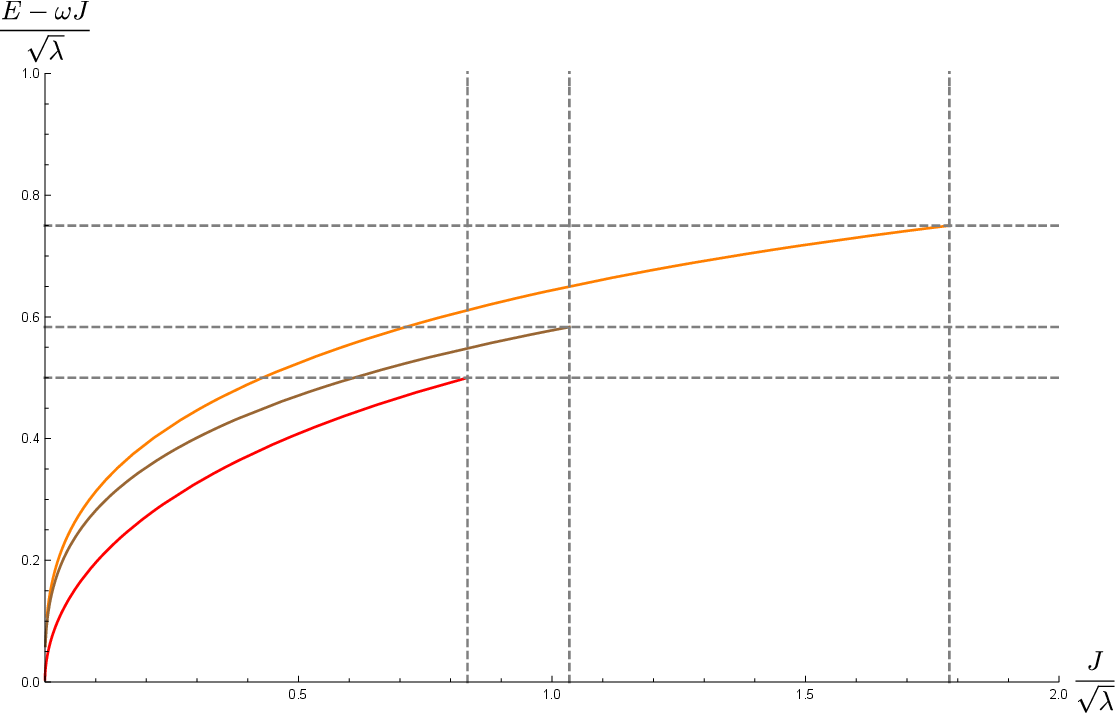}}

\ifig\loc{The plots depict the quantity $E - \omega J$ as a function of the angular momentum $J$ in units of $\sqrt{\lambda}$ in the case $n = n_0$. The brown curve is for $n = 6$ and $\gamma = -0.13497$, the orange curve is for $n = 5$ and $\gamma = -0.167391$, and the red curve is for $n = 4$ and $\gamma = -0.221068$. The horizontal dashed gray lines represent the maximum values (2.107) of $E - \omega J$. }
{\epsfxsize4.6in\epsfbox{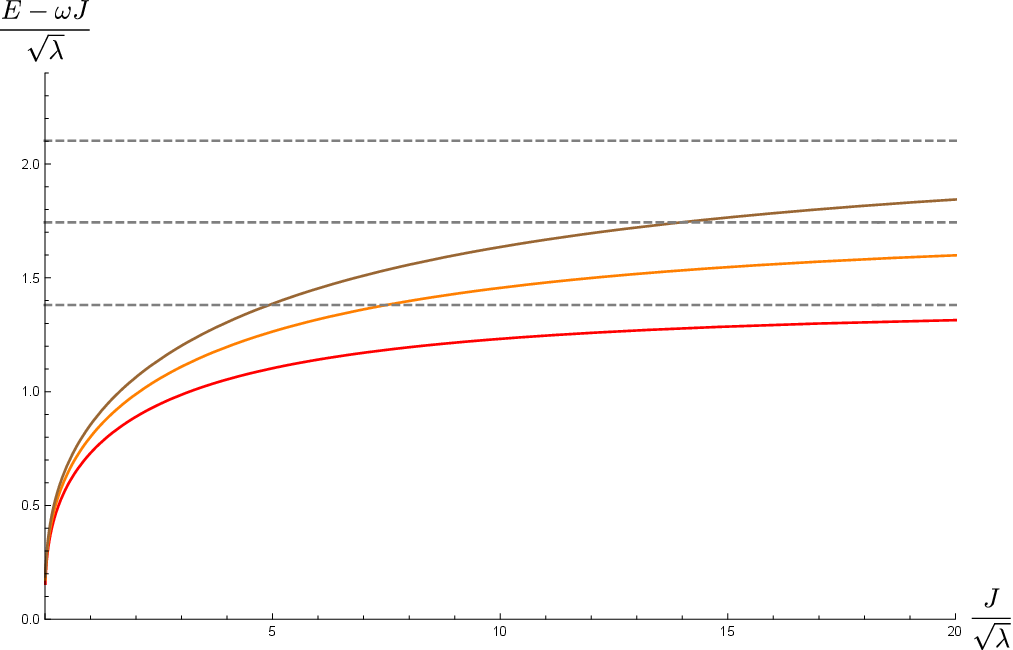}}

\ifig\loc{The plots depict the quantity $E - \omega J$ as a function of the angular momentum $J$ in units of $\sqrt{\lambda}$. The coupling $\gamma = -0.221068$ and $\mu = 4$. Thus, $n_0 = 4$. The black curve is for $n = 6$, the orange curve is for $n = 5$, the magenta curve is for $n = 4$ and the brown curve is for $n = 3$. The horizontal dashed lines represents the maximum values (2.107) and (2.109) of $E - \omega J$. The vertical dashed gray lines represent the maximum values (2.94) of $J$.}
{\epsfxsize4.5in\epsfbox{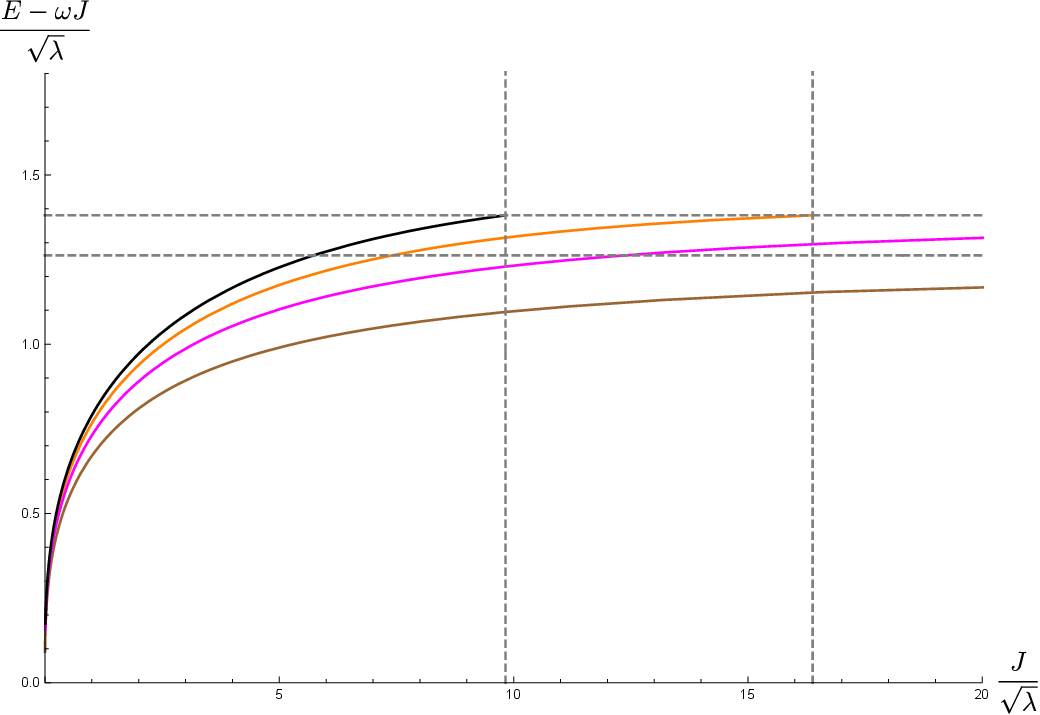}}

We also note that at $\gamma = -1$ we have $\zeta_0 = \zeta_1 = 1$. In this special case the result \integralthetaiiii\ reduces to
\eqn\special{\Delta\theta = {\pi\over 2}\left(1 - \sqrt{x_0 - 1\over x_1 - 1}\right) = {\pi \over n}.
}
This determines $x_0$ in terms of $n$ and $x_1$ or $\omega$ \cinstsevjjtheta. Note that since the number of spikes $n > 2$, $x_0 > 1$. For $n = 2$ we have $x_0 = 1$, \ie, $\tilde\theta_0 = 0$. Note also that large number of spikes implies both large $x_1$ and $x_0$ with $x_0/x_1 \approx 1$.  In fact, for large $x_0$ and $x_1$ with finite ratio $x_0/x_1$ in general we have from \integralexact,
\eqn\importrati{\eqalign{
\Delta\theta & = {1\over 1 - \gamma}\sqrt {2\gamma^2\over 1 + \gamma^2}\sqrt{x_0\over x_1}\int_{x_0}^{x_1} {dx\over x}\sqrt{x_1 - x\over x - x_0},\cr
& = {\pi \over 1 - \gamma} \sqrt {2\gamma^2\over 1 + \gamma^2}\left(1 - \sqrt{x_0\over x_1}\right) = {\pi\over n}.
}
}
We will use this result later in the section. In the case $\gamma = 0$ \integralthetaiiii\ reduces to
\eqn\twhaass{\Delta \theta = \sqrt{x_0^2 - 1\over (x_1 - 1)(x_0 + x_1)}\left[\Pi\left({x_1 - x_0\over x_1 - 1}, {x_1 - x_0\over x_1 + x_0}\right) - \Pi\left({x_1 - x_0\over x_1 + 1}, {x_1 - x_0\over x_1 + x_0}\right)\right] = {\pi \over n}.
}

An approximate and a relatively simple and useful expression in terms of (the) ordinary (hypergeometric) function can be obtained as follows. The major contribution to the integral \twhaa\ comes from the region near $\tilde\theta = \tilde\theta_0$. We first note that the denominator in the square root can be rewritten as \beforedoingGGG
\eqn\beforedoing{\eqalign{
{e^{2\phi}\sinh^22\tilde\theta - e^{2(2\phi_0 - \phi)}\sinh^22\tilde\theta_0\over \sinh(\tilde\theta - \tilde\theta_0)\sinh(\tilde\theta + \tilde\theta_0)} & =\cr
{4\left[\left(1 + \gamma^2\right)\cosh (\tilde\theta + \tilde\theta_0) - 2\gamma \cosh(\tilde\theta - \tilde\theta_0)\right]\left[\left(1 + \gamma^2\right)\cosh(\tilde\theta - \tilde\theta_0) - 2\gamma\cosh(\tilde\theta + \tilde\theta_0)\right]\over \left(1 + \gamma^2 - 2\gamma \cosh2\tilde\theta\right)\left(1 + \gamma^2 - 2\gamma \cosh2\tilde\theta_0\right)^2}.
}
}
Thus, using \beforedoing, a good approximation is
\eqn\twhbb{\eqalign{
\Delta \theta & \approx {1\over  (1 - \gamma)}\cdot {1\over\sinh\tilde\theta_1}\sqrt{\sinh(\tilde\theta_1 - \tilde\theta_0)\sinh(\tilde\theta_1 + \tilde\theta_0)\over e^{4\phi_0}\left[(1 + \gamma^2)\cosh 2\tilde\theta_0 - 2\gamma\right]\sinh 2\tilde\theta_0} \int_{\tilde\theta_0}^{\tilde\theta_1} d\tilde\theta {1\over \sqrt{\sinh(\tilde\theta - \tilde\theta_0)}}, \cr
& \approx {1\over  (1 - \gamma)}\cdot {1\over e^{2\phi_0}\sinh\tilde\theta_1}\sqrt{\sinh(\tilde\theta_1 - \tilde\theta_0)\sinh(\tilde\theta_1 + \tilde\theta_0)\over \left[(1 + \gamma^2)\cosh 2\tilde\theta_0 - 2\gamma\right] \sinh 2\tilde\theta_0} \cr
& \cdot 2\sqrt{\sinh(\tilde\theta_1 - \tilde\theta_0)}  {}_2F_1\left({1\over 4}, {1\over 2}; {5\over 4}; -\sinh^2(\tilde\theta_1 - \tilde\theta_0)\right),\cr
& \approx {2\sinh(\tilde\theta_1 - \tilde\theta_0)\over(1 - \gamma) \sinh\tilde(\theta_1)}\sqrt{\sinh(\tilde\theta_1 + \tilde\theta_0)(1 + \gamma^2 - 2\gamma\cosh2\tilde\theta_0)^2\over \sinh2\tilde\theta_0 \left[(1 + \gamma^2)\cosh 2\tilde\theta_0 - 2\gamma\right]} {}_2F_1\left({1\over 4}, {1\over 2}; {5\over 4}; -\sinh^2(\tilde\theta_1 - \tilde\theta_0)\right),
}
}
where ${}_2F_1(a, b; c; z)$ is the ordinary hypergeometric function and ${}_2F_1\left({1\over 4}, {1\over 2}; {5\over 4}; 0\right) = 1$. It obeys the transformation
\eqn\transf{\eqalign{
{}_2F_1(a, b; c; z) & =  {\Gamma(c)\Gamma(b - a)\over \Gamma(b)\Gamma(c - a)}  (-z)^{-a} {}_2F_1\left(a, a - c + 1; a - b + 1; {1\over z}\right) \cr
& + {\Gamma(c)\Gamma(a - b)\over \Gamma(a)\Gamma(c - b)}(-z)^{-b} {}_2F_1\left(b, b - c + 1; b - a + 1; {1\over z}\right),
}
}
where $|{\rm arg}(1 - z)| < \pi$. The expression \twhbb\ is relatively simple and preferable to discuss the essential characteristics of the angle difference $\Delta\theta$ as we demonstrate next.  

In the limit where $\tilde\theta_1 - \tilde\theta_0 \gg \tilde\theta_0 \gg 1$ and $\gamma \neq 0$ we get using the transformation \transf 
\eqn\twhcc{\eqalign{
\Delta\theta &\approx {4\over (1 - \gamma)}\cdot\sqrt{2\gamma^2 \over 1 +\gamma^2}\left({\Gamma(1/4)\Gamma(5/4)\over \Gamma(1/2)\sqrt{2}} + {\Gamma(-1/4)\Gamma(5/4)\over \Gamma(1/4)\Gamma(3/4)}e^{-(\tilde\theta_1 - \tilde\theta_0)/2}\right),\cr
&\approx {4\over \mu}\cdot \left({\Gamma(1/4)\Gamma(5/4)\over \Gamma(1/2)\sqrt{2}} + {\Gamma(-1/4)\Gamma(5/4)\over \Gamma(1/4)\Gamma(3/4)}e^{-(\tilde\theta_1 - \tilde\theta_0)/2}\right), \quad \mu = (1 - \gamma)\cdot\sqrt{ 1 +\gamma^2 \over 2\gamma^2 }\cr
&= {\pi\over n}. 
}
}
This implies for the string to reside entirely in the region near the boundary, \ie, the asymptotically linear dilaton region, $n$ must be greater than a minimum value $n_0$. $n_0$ depends only on the deformation coupling $\gamma$. We also note that the coefficient of the exponential term is negative. Thus, for larger $n$ values, \ie, $n > n_0$, the difference $\tilde\theta_1 - \tilde\theta_0$ must be large but finite. This ensures that \twhcc\ has a solution and thus the string confines near the boundary. In fact we will show shortly that in general $n_0 = \left \lceil{\mu}\right \rceil $ where $\left \lceil{}\right \rceil$ denotes the ceiling function.\foot{The ceiling function $\left \lceil{\xi}\right \rceil$ gives the integer greater than or equal to $\xi$.} In the case the difference $\tilde\theta_1 - \tilde\theta_0 \ll 1$ we get
\eqn\diffsmall{\Delta\theta \approx {4\over 1 - \gamma}\sqrt{2\gamma^2\over 1 + \gamma^2}\left(\tilde\theta_1 - \tilde\theta_0\right) + {\cal O}\left(\tilde\theta_1 - \tilde\theta_0\right)^2 = {\pi\over n}.
}
Note \diffsmall\ is in agreement with \importrati. In general it is clear from \twhcc\ and \diffsmall\ that large $\tilde\theta_0$ (and thus $\tilde\theta_1$ is large) implies large $n$, \ie, $n \gg n_0$.

For $\gamma = 0$ and $\tilde\theta_1 - \tilde\theta_0 \gg \tilde\theta_0 \gg 1$ we find
\eqn\twhdd{\eqalign{
\Delta\theta & \approx 2\sqrt{2}e^{-2\tilde\theta_0}\left({\Gamma(1/4)\Gamma(5/4)\over \Gamma(1/2)} \sqrt{2}+ {\Gamma(-1/4)\Gamma(5/4)\over \Gamma(1/4)\Gamma(3/4)}2e^{-(\tilde\theta_1 - \tilde\theta_0)/2}\right),\cr
&= {\pi\over n} .
}
}
We note that there is an extra multiplicative factor that decays exponentially with $\tilde\theta_0$. Thus, also in the case $\gamma = 0$ large twist or number of cusps, \ie, $n \gg 2$, implies large $\tilde\theta_0$ (and therefore $\tilde\theta_1$ is large.). 

In the limit where $\tilde\theta_1 \gg \tilde\theta_0$ and large but finite $\tilde\theta_0$ we get
\eqn\twhccc{\eqalign{
\Delta\theta &\approx {2e^{-\tilde\theta_0}\over  (1 - \gamma)}{1 + \gamma^2 - \gamma e^{2\tilde\theta_0} \over \sqrt{{(1 + \gamma^2)\over 2}e^{2\tilde\theta_0} - 2\gamma}}\left({\Gamma(1/4)\Gamma(5/4)\over \Gamma(1/2)} \sqrt{2} + {\Gamma(-1/4)\Gamma(5/4)\over \Gamma(1/4)\Gamma(3/4)}2 e^{-(\tilde\theta_1 - \tilde\theta_0)/2}\right),\cr
&= {\pi\over n}. 
}
}

In the context of CFT and string theory on AdS equivalence the spin of a (higher twist) operators is given (in global coordinates) by the angular momentum of a rigidly rotating cusped closed string. From \iii\ we have with $\kappa = 1$ the expression for the angular momentum
\eqn\twhdd{
J =  2n\cdot {\sqrt{\lambda}\over 2\pi} \int_{\tilde\theta_0}^{\tilde\theta_1} {\omega \sinh^2\tilde\theta\over \sinh (2\tilde\theta)}\sqrt{e^{2\phi}\sinh^2 (2\tilde\theta) - e^{2(2\phi_0 -\phi)}\sinh^2 (2\tilde\theta_0)\over \cosh^2\tilde\theta - \omega^2 \sinh^2 \tilde\theta}d\tilde\theta.
}
We rewrite the above integral in the following form
\eqn\jintegral{
J  = {n\sqrt{\lambda}\omega\over 4\pi}\sqrt{\zeta_0(2x_0 + \zeta_0) + 1\over -2\gamma(\zeta_0 + x_0)^2}\sqrt{x_1 - 1}\int_{x_0}^{x_1}{dx\over x+ 1}\sqrt{(x - x_0)(x + \zeta_1)\over (x_1 - x)(x + \zeta_0)},
}
where the variables $x_0, x_1, \zeta_0$ and $\zeta_1$ are given above in \cinstsevjjtheta. The integral can be expressed in terms of elliptic functions. We get
\eqn\jintegralsolve{\eqalign{
J & = {n\sqrt{\lambda}\omega\over 4\pi}\sqrt{\zeta_0(2x_0 + \zeta_0) + 1\over -2\gamma(\zeta_0 + x_0)^2}\sqrt{x_1 - 1}\cr
&\cdot \frac{2}{\left(\zeta _0-1\right) \left(x_1+1\right) \sqrt{\left(\zeta _0+x_0\right) \left(\zeta _1+x_1\right)}}\left[-\left(\zeta _0-\zeta _1\right) \left(x_1+1\right) \left(\zeta _0+x_0\right) K\left(\frac{\left(x_1-x_0\right) \left(\zeta _0-\zeta _1\right)}{\left(x_0+\zeta _0\right) \left(x_1+\zeta _1\right)}\right)+\right. \cr
& \left(\zeta _0-1\right) \left(x_1+1\right) \left(\zeta _0+x_1\right) \Pi \left(\frac{x_0-x_1}{x_0+\zeta _0}, \frac{\left(x_1-x_0\right) \left(\zeta _0-\zeta _1\right)}{\left(x_0+\zeta _0\right) \left(x_1+\zeta _1\right)}\right)- \cr
&\left. \left(\zeta _1-1\right) \left(x_0+1\right) \left(\zeta _0+x_1\right) \Pi \left(\frac{\left(x_1-x_0\right) \left(\zeta _0-1\right)}{\left(x_1+1\right) \left(x_0+\zeta _0\right)}, \frac{\left(x_1-x_0\right) \left(\zeta _0-\zeta _1\right)}{\left(x_0+\zeta _0\right) \left(x_1+\zeta _1\right)}\right)\right].
}
}

In the special case $\gamma = -1$ we have $\zeta_0 = \zeta_1 = 1$ \cinstsevjjtheta, and the angular momentum $J$ is given by
\eqn\specialJ{J = \frac{n \sqrt{\lambda }  \omega}{4}   \sqrt{\frac{x_1 -1}{x_0 +1}} \left(1 - \sqrt{\frac{x_0 +1}{x_1 +1}}\right).
}
We note that for $x_0 = \cosh\tilde\theta_0 = 1$, \ie, $n = 2$, the angular momentum \specialJ\ agrees with the result \iiiiiii. We can rewrite \specialJ\ in terms of $\omega$ and $n$ using the result \special\ and the definitions \cinstsevjjtheta. This gives
\eqn\specialangulaRTr{J = {n\sqrt{\lambda}\over 4}\left[{\omega\over \sqrt{\omega^2 - 1 + \left(1 - {2\over n}\right)^2}} - 1\right], \quad n \geq 2.
}
 It follows that in the case $n > 2$ the angular momentum $J$ is bounded from above. The bound is obtained by setting $\omega = 1$ or equivalently, as we will show shortly, by taking the large $x_1$ limit. For a given $n$ the maximum value it can take is given by
\eqn\specialangular{J = {\sqrt{\lambda}\over 2}\left(n\over n - 2\right),
}
We note that for $n > 2$ it is finite. We also note that the maximum $J$ value, \ie, \specialangular, is bounded from below by $\sqrt{\lambda}/2$ which corresponds to taking $n$ large or equivalently taking both $x_0$ and $x_1$ large, see \special\ and \twhcc. In general in the case in which both $x_0$ and $x_1$ are large but with finite ratio $x_0/x_1$ we get from \jintegral
\eqn\importjj{\eqalign{
J &= {n\sqrt{\lambda}\over 4\pi}\sqrt{1 + \gamma^2 \over 2\gamma^2}\sqrt{x_1\over x_0}\int_{x_0}^{x_1}{dx\over x}\sqrt{x - x_0\over x_1 - x},\cr
& = {n\sqrt{\lambda}\over 4}\sqrt{1 + \gamma^2\over 2\gamma^2}\left(\sqrt{x_1\over x_0} - 1\right).
}
}
We will comment on this further later in this section.

In the case $\gamma = 0$ we rewrite \jintegral\ as
\eqn\twhdds{
J =  {n\omega\sqrt{\lambda}\over 2\pi}\sqrt{x_1}\int_{x_0}^{x_1}{dx\over x + 1}\sqrt{(x- x_0)(x + x_0 + 1)\over (x_1 - x)},
}
here $x_0 = \sinh^2\tilde\theta_0$ and $x_1 = \sinh^2\tilde\theta_1$. We find
\eqn\twhddss{\eqalign{
J & =  {n\omega\sqrt{\lambda}\over \pi}\sqrt{x_1} \left[\sqrt{x_1 + x_0 + 1}E\left({x_1 - x_0\over x_1 + x_0 + 1}\right)-\right. \cr
&  {x_0 + 1\over \sqrt{x_1 + x_0 + 1}}K\left({x_1 - x_0\over x_1 + x_0 + 1}\right) - \cr
&\left.{x_0(x_0 + 1)\over (x_1 + 1)\sqrt{x_1 + x_0 + 1}}\Pi\left({x_1 - x_0\over x_1 + 1},{x_1 - x_0\over x_1 + x_0 + 1}\right)\right].
}
}
In the limit $x_0/x_1 \ll 1$, \ie, $\omega \approx 1$ this gives, to leading order, 
\eqn\twhddss{
J = {n\sqrt{\lambda}\over \pi}x_1, \quad x_1 = {1\over \omega^2 - 1} \gg \sqrt{n}.
}

A relatively simple and useful expression can be obtained in a similar way as in the previous subsection. The dominant contribution to the integral \jintegral\ comes from near $\tilde\theta = \tilde\theta_1$. Thus, a good approximation is
\eqn\twhddd{\eqalign{
J & \approx {2n\over 2\pi}\cdot \omega\sqrt{\lambda}\cdot {e^{2\phi_0}\sinh2\tilde\theta_0\over 2e^{2\phi_1}\cosh^2\tilde\theta_1} \int_{\tilde\theta_0}^{\tilde\theta_1} {d\tilde\theta} {\tilde\theta'}, \cr
& \approx {2n\over 2\pi}\cdot \omega\sqrt{\lambda}\cdot {\tanh\tilde\theta_1\over 2 e^{\phi_1}}\sqrt{e^{4\phi_1}\sinh^2 2\tilde\theta_1 - e^{4\phi_0}\sinh^2 2\tilde\theta_0} \int_{\tilde\theta_0}^{\tilde\theta_1} {d\tilde\theta\over \sqrt{\cosh^2\tilde\theta - \omega^2\sinh^2\tilde\theta}}, \cr
& \approx {2n\over 2\pi}\cdot \omega\sqrt{\lambda}\cdot {\tanh\tilde\theta_1\over 2 e^{\phi_1}}\sqrt{e^{4\phi_1}\sinh^2 2\tilde\theta_1 - e^{4\phi_0}\sinh^2 2\tilde\theta_0}\cdot {\sqrt{\tanh\tilde\theta_1\over 2}} \int_{\tilde\theta_0}^{\tilde\theta_1} {d\tilde\theta\over \sqrt{\sinh(\tilde\theta_1 - \tilde\theta)}},\cr
& \approx {n\omega\sqrt{\lambda}\over \sqrt{2}\pi}  {\tanh^{3/2}\tilde\theta_1\over e^{\phi_1}}\sqrt{\sinh(\tilde\theta_1 - \tilde\theta_0)(e^{4\phi_1}\sinh^2 2\tilde\theta_1 - e^{4\phi_0}\sinh^2 2\tilde\theta_0)}\cdot {}_2F_1\left({1\over 4}, {1\over 2}; {5\over 4}; -\sinh^2(\tilde\theta_1 - \tilde\theta_0)\right).
}
}
In the limit where $\tilde\theta_1 - \tilde\theta_0 \gg \tilde\theta_0 \gg 1$ and $\gamma \neq 0$ we get, using \cinstsevjjtheta\ and \beforedoing,
\eqn\twhee{\eqalign{
J &\approx {n\over 2\pi}\sqrt{\lambda\over \pi}{\sqrt{1 + \gamma^2}\over -\gamma}\Gamma(1/4)\Gamma(5/4)e^{\tilde\theta_1-\tilde\theta_0} + {\cal O}\left(e^{(\tilde\theta_1- \tilde\theta_0)/2}\right),\cr
& \approx {\sqrt{2}\Gamma^2(1/4)\over 2\pi\Gamma(1/2)} \cdot {n\sqrt{\lambda}\over 4}\sqrt{1 + \gamma^2\over 2\gamma^2}\cdot e^{\tilde\theta_1 - \tilde\theta_0} + {\cal O}\left(e^{(\tilde\theta_1- \tilde\theta_0)/2}\right)  \approx  {\sqrt{2}\Gamma^2(1/4)\over 2\pi\sqrt{\pi}}{n\sqrt{\lambda}\over 4}\sqrt{1 + \gamma^2 \over 2\gamma^2}e^{\tilde\theta_1} + {\cal O}\left(e^{\tilde\theta_1/2}\right). 
}
}
Thus, in general, large $\tilde\theta_1 - \tilde\theta_0$ implies larger angular momentum. As we saw earlier in \twhcc,\ this also implies $n \approx n_0$. As we will demonstrate shortly, for larger $n$, \ie, $n > n_0$, the angular momentum is bounded from above. We can actually already observe this phenomenon from \twhee\ since for large but finite difference $\tilde\theta_1 - \tilde\theta_0$ and thus $n > n_0$ \twhcc, the angular momentum $J$ is large but finite. For small difference, \ie, $\tilde\theta_1 - \tilde\theta_0 \ll 1$ we instead get
\eqn\smalldiffJ{J \approx {n\sqrt{\lambda}\over \pi}\sqrt{1 + \gamma^2\over 2\gamma^2}\left(\tilde\theta_1 - \tilde\theta_0\right) + {\cal O}\left(\tilde\theta_1 - \tilde\theta_0\right)^2.
}
Note \smalldiffJ\ is in agreement with \importjj. 

In the case $\gamma = 0$ and $\tilde\theta_1 - \tilde\theta_0 \gg \tilde\theta_0 \gg 1$ we find
\eqn\twheee{\eqalign{
J &\approx {n\over 2\pi}\sqrt{\lambda\over 2\pi}\Gamma(1/4)\Gamma(5/4)e^{2\tilde\theta_1} + {\cal O}\left(e^{(\tilde\theta_1+ \tilde\theta_0)}\right),\cr
&  \approx {\Gamma^2(1/4)\over 8\pi\sqrt{2\pi}} \cdot n\sqrt{\lambda}e^{2\tilde\theta_1} + {\cal O}\left(e^{(\tilde\theta_1+ \tilde\theta_0)}\right) \approx {\Gamma^2(1/4)\over 8\pi\sqrt{\pi}} \cdot n\sqrt{\lambda}e^{2\tilde\theta_1} + {\cal O}\left(e^{\tilde\theta_1}\right).
}
}
Thus, large $\tilde\theta_1$ implies large $J$.

In the limit where $\tilde\theta_1 \gg \tilde\theta_0$ and (large but) finite $\tilde\theta_0$ we get
\eqn\twhff{\eqalign{
J &\approx {n\over 2\pi}\sqrt{\lambda\over \pi}\sqrt{1 + \gamma^2}\Gamma(1/4)\Gamma(5/4){e^{\tilde\theta_1 + \tilde\theta_0}\over 1 + \gamma^2 - 2\gamma\cosh(2\tilde\theta_0)} + {\cal O}\left(e^{(\tilde\theta_1+ \tilde\theta_0)/2}e^{\tilde\theta_0}\right), \quad \gamma \neq 0,\cr
& \approx {\Gamma^2(1/4)\over 8\pi\sqrt{\pi}} \cdot  n{\sqrt{\lambda(1 + \gamma^2)}e^{\tilde\theta_1}\over 1 + \gamma^2 - 2\gamma\cosh(2\tilde\theta_0)} + {\cal O}\left(e^{\tilde\theta_1/2}\right), \quad \gamma \neq 0. 
}
}

In perturbation theory in the coupling $\gamma$ we also have
\eqn\nnnv{\eqalign{
J & =  2n\cdot {\sqrt{\lambda}\over 2\pi} \int_{\tilde\theta_0}^{\tilde\theta_1} {\omega \sinh^2\tilde\theta\over \sinh (2\tilde\theta)}\sqrt{e^{2\phi}\sinh^2 (2\tilde\theta) - e^{2(2\phi_0 -\phi)}\sinh^2 (2\tilde\theta_0)\over \cosh^2\tilde\theta - \omega^2 \sinh^2 \tilde\theta}d\tilde\theta,\cr
& = {n\omega\sqrt{\lambda}\over \pi} \int_{\tilde\theta_0}^{\tilde\theta_1} d\tilde\theta { \sinh^2\tilde\theta\over \sinh 2\tilde\theta}\sqrt{\sinh^2 2\tilde\theta - \sinh^2 2\tilde\theta_0\over \cosh^2\tilde\theta - \omega^2 \sinh^2 \tilde\theta}\cr
&\cdot \left[1 + \gamma\left({2\sinh^22\tilde\theta\over \cosh 2\tilde\theta + \cosh 2\tilde\theta_0} + 2\cosh 2\tilde\theta_0 - \cosh 2\tilde\theta\right) + {\cal O}\left(\gamma^2\right)\right],\cr
& = {n\omega\sqrt{\lambda}\over \pi}\sqrt{x_1}\left\{{1\over 2}I_2 + \gamma\left[2I_3 + (2x_0 + 1)I_2 - {1\over 2}(2I_1 - I_2)\right] + {\cal O}\left(\gamma^2\right)\right\},\cr
& = {n\omega\sqrt{\lambda}\over \pi}\sqrt{x_1}\left\{{1\over 2}I_2 + \gamma\left[2I_3 - I_1 + \left(2x_0 + {3\over 2}\right)I_2\right] + {\cal O}\left(\gamma^2\right)\right\},
}
}
here  $x_0 = \sinh^2\tilde\theta_0$ and $x_1 = \sinh^2\tilde\theta_1$. The integrals $I_1, I_2$ and $I_3$ are given in terms of elliptic functions in appendix A. In the limit $x_0/x_1 = \xi \ll 1$, \ie, $\omega \approx 1$, we have $I_1 = 4x_1^{3/2}/3, I_2 = 2x_1^{1/2}$ and $I_3  = 4x_1^{3/2}/3$. Thus, in this limit we get
\eqn\pertubjlimit{\eqalign{
J & = {n\sqrt{\lambda}\over \pi}x_1\left[ 1 + {4\over 3} x_1\gamma + {\cal O}\left(\gamma^2\right)\right], \quad x_1 = {1\over \omega^2 - 1}, \quad \omega = 1 + 2\eta, \quad \eta \ll 1,\cr
& = {n\sqrt{\lambda}\over \pi}\cdot {1\over 4\eta}\left[ 1 + {4\over 3} \cdot {1\over 4\eta}\gamma + {\cal O}\left(\gamma^2\right)\right].
}
}
We note that with $n = 2$ it gives \twjfff. We also note that the radial distance of the spike is 
\eqn\lenn{\tilde\theta_1 =  {1\over 2} \left[\ln\left(4\pi {J\over n\sqrt{\lambda}}\right) - {4\pi \over 3}{J\over n \sqrt{\lambda}}\gamma + {\cal O}\left({J^2\over \lambda}\gamma^2\right)\right].
}
Thus, the deformation washes the boundary away \AsratMHB.

In general, the angular momentum is bounded from above for $n > n_0 = \lceil{\mu} \rceil$ where $\lceil{\mu}\rceil$ gives the integer greater than or equal to $\mu$. The maximum value is given in general by 
\eqn\maxjjj{\eqalign{
J & = {\sqrt{\lambda}\over 4} {(1 - \gamma)(1 + \gamma^2)\over 2\gamma^2}\left({n\over n - (1 - \gamma)\sqrt{{1 + \gamma^2 \over 2\gamma^2}}}\right),\cr
 & = {\sqrt{\lambda}\over 4(1 - \gamma)}\left({n\mu^2\over n - \mu}\right), \quad \mu := (1 - \gamma)\sqrt{1 + \gamma^2\over 2\gamma^2}.
}
}
This is obtained using \importrati\ and \importjj. We note that $\mu(\gamma) = \mu({1\over \gamma})$. Therefore, $\gamma = -1$ is a self dual point. Away from $\gamma = -1$ $\mu$ increases. For $\gamma = -1$ we have $\mu = 2$ and therefore \maxjjj\ gives the earlier result \specialangular. Thus, in the case $\gamma = -1$, the angular momentum is bounded for any $n$, see Fig. 4. For $n \leq n_0$ the angular momentum is not bounded, see Fig. 6 and Fig. 7. For a given $\gamma$ the bound decreases with increasing $n$. The maximum value \maxjjj\ is bounded from below by $\sqrt{\lambda} \mu^2/(4(1-\gamma))$ which only depends on $\gamma$. For a given $n$ the bound \maxjjj\ also decreases with increasing the norm of $\gamma$, see Fig. 5. 

We next consider the combination $E - \omega J$ which gives at $\omega = 1$ and $\gamma = 0$ the anomalous dimensions of the boundary operators with twist $n > 2$. We have from \kkkk\ with $\kappa = 1$
\eqn\twhfff{\eqalign{
E - \omega J & = 2n{\sqrt{\lambda}\over 2\pi}\int_{\tilde\theta_0}^{\tilde\theta_1} {d\tilde\theta\over {\tilde\theta'}}{\cal L}, \quad {\cal L} = {\kappa_\gamma\over 2}\cdot {e^{4\phi}\sinh^22\tilde\theta\over e^{2\phi_0}\sinh 2\tilde\theta_0}, \cr
& = {n\sqrt{\lambda}\over \pi}\int_{\tilde\theta_0}^{\tilde\theta_1}e^{2\phi}\sinh 2\tilde\theta \sqrt{\cosh^2\tilde\theta - \omega^2 \sinh^2\tilde\theta\over e^{2\phi}\sinh^2 2\tilde\theta - e^{2(2\phi_0 -\phi)}\sinh^2 2\tilde\theta_0}d\tilde\theta.
}
}
We rewrite the integral as
\eqn\emjjgeneral{\eqalign{
E - \omega J & = {n\sqrt{\lambda}\over \pi}{1\over 2}{\zeta _0+ x_0\over \sqrt{(x_1 - 1)[(1 + \gamma^2)(2x_0 + \zeta_0)-2\gamma]}}\int_{x_0}^{x_1}\sqrt{x_1 - x\over (x - x_0)(x + \zeta_0)(x + \zeta_1)}dx,
}
}
where the variables $x_0, x_1, \zeta_0$ and $\zeta_1$ are given in \cinstsevjjtheta. The integral can be rewritten using elliptic functions. We get
\eqn\emjjgeneral{\eqalign{
E - \omega J  & = {n\sqrt{\lambda}\over \pi}{1\over 2}{\zeta _0+ x_0\over \sqrt{(x_1 - 1)[(1 + \gamma^2)(2x_0 + \zeta_0)-2\gamma]}}\cr
&\cdot  {2(\zeta_0 + x_1)\over \sqrt{(x_1 + \zeta_1)(x_0 + \zeta_0)}}\left[K\left({(x_1 - x_0)(\zeta_0 - \zeta_1)\over (x_1 + \zeta_1)(x_0 + \zeta_0)}\right) - \Pi\left(-{x_1 - x_0\over x_0 + \zeta_0}, {(x_1 - x_0)(\zeta_0 - \zeta_1)\over (x_1 + \zeta_1)(x_0 + \zeta_0)}\right)\right].
}
}
In the special case $\zeta_0 = \zeta_1 = 1$ we have the simple exact result
\eqn\specialemj{E - wJ = {n\sqrt{\lambda}\over 4}\sqrt{{x_0 + 1\over x_1 - 1}}\left(\sqrt{x_1 + 1\over x_0 + 1} - 1\right).
}
We note that for $x_0 = \cosh\tilde\theta_0 = 1$, \ie, $n = 2$, the energy-momentum relation \specialemj\ agrees with the earlier result we obtained in \kkkkkkkkk.  We rewrite \specialemj\ using \special\ and \cinstsevjjtheta\ as
\eqn\eminusomegafj{E - \omega J = {n\sqrt{\lambda}\over 4}\left[\omega - \sqrt{\omega^2 - 1 + \left(1 - {2\over n}\right)^2}\right].
}
We note that $E - \omega J$ is bounded from above. The maximum value it can take is
\eqn\specoundenergy{E -  J = {\sqrt{\lambda}\over 2},
}
which is independent of $n$. As we will show shortly the maximum value in general depends both on the coupling $\gamma$ and the number of cusps or spikes $n$. We note that in the special case $\gamma = -1$ we get using \specialJ\ the result
\eqn\specialemjj{E = \left[\omega + \sqrt{\omega^2 - 1 + \left(1 - {2\over n}\right)^2}\right]J, \quad n \geq 2.
}
This gives the simple exact Regge relation
\eqn\specRegge{m^2 = 4\left({n - 1\over n^2}\right)\cdot (l + n)l, \quad n \geq  2,
}
where $m$ and $l$ are given in \renormal. In flat spacetime we have \Kruczenski
\eqn\flatRegge{m^2 = 8\left({n - 1\over n}\right)l, \quad n \geq 2.
}
Thus, in the case $n > 2$ the result \specRegge\ in the limit $l \ll n$ is qualitatively similar but different to that of \flatRegge. Since the coupling $\gamma$ is non zero in \specRegge, the two results are not in general expected to be identical.

In the case $\gamma = 0$ we rewrite the integral \twhfff\ as
\eqn\twhhhb{\eqalign{
E - \omega J & = {n\over 2\pi}\sqrt{\lambda\over x_1}\int_{x_0}^{x_1}dx \sqrt{x_1 - x\over (x - x_0)(x + x_0 + 1)},\cr
& = {n\over 2\pi}\sqrt{\lambda\over x_1}\cdot 2\sqrt{x_1 + x_0 + 1}\left[K\left({x_1 - x_0\over x_1 + x_0 + 1}\right) - E\left({x_1 - x_0\over x_1 + x_0 + 1}\right)\right],
}
}
here $x_0 = \sinh^2\tilde\theta_0$ and $x_1 = \sinh^2\tilde\theta_1$. In the limit $x_0/x_1 \ll 1$ we get
\eqn\twhhhbxz{E - \omega J = -{n\sqrt{\lambda}\over 2\pi}\ln\xi+ {\cal O}\left(\xi^2\ln\xi\right), \quad \xi = {x_0\over x_1}, \quad x_1 = {1\over \omega^2 - 1}.
}

We now determine the coupling $\gamma$ dependence of the maximum value \specoundenergy\ of $E - \omega J$. The dominant contribution to the integral \twhfff\ in the case $\gamma \neq 0$ comes from the region near $\tilde\theta = \tilde\theta_0$. Thus, a good approximation is obtained by replacing ${\cal L}(\tilde\theta)$ by ${\cal L}(\tilde\theta_0)$. This gives 
\eqn\twhggg{\eqalign{
E - \omega J & \approx 2n\cdot {\sqrt{\lambda}\over 2\pi}\cdot (1 - \gamma)\cdot {1\over 2} e^{2\phi_0}\sinh 2\tilde\theta_0\cdot \int_{\tilde\theta_0}^{\tilde\theta_1} {d\tilde\theta\over {\tilde\theta'}},\cr
 & \approx (1 - \gamma)\cdot {\sqrt{\lambda}\over 2}e^{2\phi_0}\sinh 2\tilde\theta_0, \cr
 & \approx {\sqrt{\lambda}\over 2} {(1 - \gamma)\sinh 2\tilde\theta_0\over 1 + \gamma^2 - 2\gamma \cosh 2\tilde\theta_0}.
}
}
In the limit of large angular momentum, \ie, $\omega \approx 1$ and/or large number of cusps or spikes, \ie, $n \gg 2$, to be precise $n \geq n_0$, we then find
\eqn\twhgggg{E -  J =  {\sqrt{\lambda}\over 4}\cdot {(1 - \gamma) \over - \gamma }.
}
The maximum value \twhgggg\ can be also obtained by studying the large $x_0$ and $x_1$ limit of \emjjgeneral. Taking this limit but keeping their ratio $x_0/x_1$ finite we get
\eqn\specimaxemj{\eqalign{
E - \omega J & = {n\sqrt{\lambda}\over \pi}{1\over 2}\sqrt{x_0\over 2(1 + \gamma^2)x_1}\int_{x_0}^{x_1}{dx\over x}\sqrt{x_1 - x\over x - x_0},\cr
& = {\sqrt{\lambda}\over 2\sqrt{2}}{n\over\sqrt{1 + \gamma^2}}\left(1 -\sqrt{x_0\over x_1}\right).
}
}
Upon using \importrati\ in \specimaxemj\ we get the exact result \twhgggg.

In the case $\omega \approx 1$ but finite $x_0$ or $\tilde\theta_0$, \ie, $n \leq n_0$, we get the maximum value
\eqn\maxefbe{
E - J = {n\sqrt{\lambda}\over \pi}{\sqrt{(1 - \gamma)^2 - 2\gamma (x_0 - 1)}\over \sqrt{(1 - \gamma)^4 - 4\gamma(1+\gamma^2)(x_0 - 1)}}K\left({(1 - \gamma)^2(1 + \gamma)^2}\over (1 - \gamma )^4 - 4\gamma(1 + \gamma^2)(x_0 - 1)\right),
}
where $x_0 = \cosh 2\tilde\theta_0$ is determined using the angle difference \integralthetaiiii. In the case $n = 2$ or $x_0 = \cosh 2\tilde\theta_0 = 1$ it reduces to \twtomegaone. In general \maxefbe\ depends both on $n$ and $\gamma$, see Fig. 7. In the special $\gamma = -1$ case however it is independent of $n$ since $n_0 = 2$. In general for $n > n_0$ the maximum value \twhgggg\ is independent of $n$, see Fig. 4.

To examine the behavior of $E - \omega J$ for small $\gamma$ we expand \twhfff\ around $\gamma = 0$.  In perturbation theory in the coupling $\gamma$ we find
\eqn\twhhh{\eqalign{
E - \omega J & = 2n{\sqrt{\lambda}\over 2\pi}\int_{\tilde\theta_0}^{\tilde\theta_1}e^{2\phi}\sinh 2\tilde\theta \sqrt{\cosh^2\tilde\theta - \omega^2 \sinh^2\tilde\theta\over e^{2\phi}\sinh^2 2\tilde\theta - e^{2(2\phi_0 -\phi)}\sinh^2 2\tilde\theta_0}d\tilde\theta, \cr
& = 2n{\sqrt{\lambda}\over 2\pi}\int_{\tilde\theta_0}^{\tilde\theta_1}d\tilde\theta\sinh 2\tilde\theta \sqrt{\cosh^2\tilde\theta - \omega^2 \sinh^2\tilde\theta\over \sinh^2 2\tilde\theta - \sinh^2 2\tilde\theta_0}\cdot\cr
& \left[1 + \gamma\left(-2\cosh 2\tilde\theta_0 + 3\cosh 2\tilde\theta - {2\sinh^2 2\tilde\theta\over \cosh 2\tilde\theta + \cosh 2\tilde\theta_0}\right) + {\cal O}(\gamma^2)\right],\cr
& = {n\over \pi}\sqrt{\lambda\over x_1}\left\{{1\over 2}I_4 + \gamma\left[-(2x_0 + 1)I_4 + {3\over 2}I_5 - I_6\right] + {\cal O}(\gamma^2)\right\},
}
}
here $x_0 = \sinh^2\tilde\theta_0, x_1 = \sinh^2\tilde\theta_1$. The integrals $I_4, I_5$ and $I_6$ are given in appendix A in terms of elliptic functions. 

In the limit $x_0/x_1 \ll 1$ we get $I_4 = \sqrt{x_1}\ln x_1$, $I_5 = 4x_1^{3/2}/3$, and $I_6 = 4x_1^{3/2}/3$. Thus,
\eqn\twhhhbxz{\eqalign{E - \omega J & = {n\sqrt{\lambda}\over \pi}\left[{1\over 2}\ln x_1 + {2\over 3}x_1 \gamma + {\cal O}\left(\gamma^2\right)\right], \quad x_1 = {1\over \omega^2 - 1}, \quad \omega = 1 + 2\eta, \quad \eta \ll 1\cr
& =  {n\sqrt{\lambda}\over \pi}\left[{1\over 2}\ln \left({1\over \eta}\right) + {2\over 3}\cdot {1\over 4\eta} \gamma + {\cal O}\left(\gamma^2\right)\right].
}
}
We note that for $n = 2$ it reduces to the second equation in \twemjfffx. Using the result \pertubjlimit\ the correction to $E - J$ is
\eqn\correctiondi{E - J = {n\sqrt{\lambda}\over 2\pi}\left[\ln\left({4 \pi \over n} {J\over \sqrt{\lambda}}\right) + {4\pi \over 3n}{J\over \sqrt{\lambda}}\gamma +  {\cal O}(\gamma^2)\right]. 
}
We also note here the first non vanishing correction starts at order $\gamma$. At zeroth order we have the anomalous dimension of an operator with twist $n$. The first correction is independent of (the twist) $n$, and thus similar for all the spinning strings. Therefore, the quantity $E - J$ do not scale linearly with $n$. See also \refs{\Callan,\  \Belitsky} for a related discussion.

Using \twhgggg\ and \maxjjj\ we have in the limit $\omega \approx 1$ the linear relation between the energy and angular momentum,
\eqn\linregge{E = \left[\left({n - \mu\over  n}\right)\left({-2\gamma\over 1 + \gamma^2}\right) + 1\right]J, \quad n > n_0.
}
This is in agreement with \specialemjj\ in the special case $\gamma = -1$ (and $\omega = 1$). In the large $n$ limit, \ie, $n \gg n_0$, this gives
\eqn\linreggGHe{E = \left[{(1 - \gamma)^2\over 1 + \gamma^2}\right]J, \quad \gamma \neq 0.
}

\newsec{Cusp anomalous dimension}

An alternative and perhaps an effective approach to determine the anomalous dimensions $f(\lambda)\ln S + {\cal O}(1/S)$ (of twist two operators in gauge theories) is using Wilson loops. The approach is based on a Wilson loop containing a light-like cusp singularity. A cusp anomalous dimension arises due to the renormalization property of the cusp singularity. It gives the coefficient of the logarithmic term, \ie, $f(\lambda)$. The function $f(\lambda)$ interpolates between $\lambda$ at weak coupling and ${\sqrt{\lambda}}$ at strong coupling (up to numerical factors which in general depend on the representation in which the Wilson loop is evaluated).

The vacuum expectation of a Wilson loop over a closed integration contour $C_\delta$ containing a single light-like cusp with (minkowskian) cusp angle $\delta$ is given by \KorchemskyW
\eqn\expectwill{
\langle W_\delta \rangle \sim \exp\left[-\Gamma_{\rm cusp}(\lambda, \delta)\ln\left({\mu \over \Lambda }\right) - \cdots\right],
}
where $\mu$ is a UV cutoff and $\Lambda$ is an IR cutoff. The dots denote terms containing higher powers of $\ln(\mu/\Lambda)$. They show up starting at two loop. The coefficient $\Gamma_{\rm cusp}$ measures the cusp or soft anomalous dimension of the Wilson loop. It only exists because of the cusp, and thus it depends only on the cusp angle $\delta$. For $\delta \gg 1$ $\Gamma_{\rm cusp}$ is linear in $\delta$ for arbitrary order in perturbation \KorchemskyW. The anomalous dimension $\Gamma_{\rm cusp}$ determines the function $f(\lambda)$ \KruczenskiA.  

In the context of gauge/gravity duality, on the other hand, the vacuum expectation value of a given Wilson loop on a contour $C_\delta$ is given by the (fundamental) string worldsheet action $S$ whose boundary coincides with $C_\delta$ \MaldacenaW,
\eqn\anomal{
\langle W(C_\delta)\rangle = \exp(-S(C_\delta)).
}

In what follows, we use the equivalence \anomal\ together with \expectwill\ to compute the one loop cusp anomalous dimension and examine its relation with \twfff, \ie, the quantity $E - J$, in the case the string theory metric is asymptotically non-AdS. At $\gamma = 0$ they are equal (in the sense we below show). See \KruczenskiRT\ for a detailed discussion.

The cusp anomalous dimension depends only on the cusp angle. Thus, we consider two semi-infinite light-like lines in the boundary of ${\cal A}_3$ meeting at a point. The bulk metric in Poincar\'e coordinates is \bbbbbbv\ (written here again)
\eqn\eeee{ds^2 = d\tilde\theta^2 + f(\tilde\theta; \gamma)\left(-d\tilde\varphi^2 + d\tilde\psi^2\right), \quad f(\tilde\theta; \gamma) = {1\over 4} {e^{2\tilde\theta}\over 1 - \gamma e^{2\tilde\theta}} = {1\over 4}e^{2\tilde\theta} e^{2\phi},
}
where $\phi$ is the dilaton and $\tilde\psi$ takes value in $\IR$.

A convenient choice of parameterization of the worldsheet is
\eqn\fffff{\tilde\varphi = e^\tau \cosh\sigma, \quad \tilde\psi = e^\tau\sinh\sigma, \quad u = e^{-\tilde\theta} = e^\tau q(\gamma e^{-2\tau}, \sigma), 
}
where $q$ is some function. In general $q$ is a function of $\tau$ and $\sigma$. It obeys the boundary condition
\eqn\ggggg{\lim_{\tau \to -\infty} e^{\tau}q(\gamma e^{-2\tau}, \sigma) = 0.
}
We note that boost symmetry is now translation in $\sigma$.\foot{The metric in the variables $(\tau, \sigma, u)$ is
\eqn\ababab{ds^2 = {du^2\over u^2} + f(u;\gamma)e^{2\tau}\left(-d\tau^2 + d\sigma^2\right), \quad f(u; \gamma) = {1\over 4}{u^{-2}\over 1 - \gamma u^{-2}}.
}
Translation in $\sigma$ is a symmetry of the metric. We also note that under the transformations $\tau \to \tau + \kappa, u \to u e^{\kappa}$ and $\gamma \to \gamma e^{2\kappa}$ where $\kappa$ is a constant the metric is invariant.}
Thus, the function $q$ is independent of $\sigma$. We write
\eqn\minareafun{q(\gamma e^{-2\tau}, \sigma) = h(\gamma e^{-2\tau}),
}
where $h$ is a function of only $\tau$.

The induced metric is thus given by
\eqn\iiiiii{ds^2 = -\left[fe^{2\tau} - \left(1 + {1\over h}\partial_\tau h\right)^2\right]d\tau^2 + f e^{2\tau}d\sigma^2.
}

The area of the induced surface (or the Nambu Goto action for a fundamental open string) is 
\eqn\kkkk{\eqalign{
S & = {\sqrt{\lambda}\over 2\pi} \cdot 2 \int_{0}^{\infty}d\sigma \cdot  \int_{-\infty}^{\infty} d\tau f e^{2\tau}\sqrt{-1 + {1\over f e^{2\tau}}\left(1 + {1\over h}\partial_\tau h\right)^2},\cr
& = {\sqrt{\lambda}\over 2\pi} \cdot 2  \Lambda_\sigma \cdot \int_{-{\Lambda_\tau}}^{\Lambda_\tau} d\tau  {\sqrt{-1 + 4\left(h^2 - \gamma e^{-2\tau}\right)\left(1 + {1\over h}\partial_\tau h\right)^2}\over 4(h^2 - \gamma e^{-2\tau})},\cr
& = {\sqrt{\lambda}\over 2\pi} \cdot 2  \Lambda_\sigma \cdot \int_{-{\Lambda_\tau}}^{\Lambda_\tau} d\tau {\cal L}, \quad {\cal L} =  {\sqrt{-1 + 4\left(h^2 - \gamma e^{-2\tau}\right)\left(1 + {1\over h}\partial_\tau h\right)^2}\over 4(h^2 - \gamma e^{-2\tau})}.
}
}
where $\Lambda_\sigma$ and $\Lambda_\tau$ are large cut-offs. $2\Lambda_\sigma$ gives the cusp angle $\delta$. $\Lambda_\tau$ is related to the cutoffs $\mu$ and $\Lambda$. We will give the expressions for both $\Lambda_\sigma$ and $\Lambda_\tau$ shortly. The equation of motion for $h$ is given by Euler's equation
\eqn\lllll{\partial_\tau {\partial {\cal L}\over \partial \partial_\tau h} - {\partial {\cal L}\over \partial h} = 0.
}

In perturbation theory we get
\eqn\iiiiiiii{\eqalign{
h&={1\over \sqrt{2}}\left[1 + \ln\left(1 - 3\gamma e^{-2\tau} -{30\over 7}\gamma^2 e^{-4\tau} + {\cal O}\left(\gamma^3\right)\right)\right], \cr
&= {1\over \sqrt{2}}\left[1 - 3\gamma e^{-2\tau} -{123\over 14}\gamma^2 e^{-4\tau} +  {\cal O}\left(\gamma^3\right)\right].
}
}
Thus,
\eqn\rhorho{u = e^{-\tilde\theta} = {e^{\tau}\over \sqrt{2}}\left[1 - 3\gamma e^{-2\tau} -{123\over 14}\gamma^2 e^{-4\tau} +  {\cal O}\left(\gamma^3\right)\right].
}
Up to and including order $\gamma^2$ we then get
\eqn\divergences{\eqalign{
 \int_{\epsilon}^{L} {du\over u} & = \ln\left({L\over \epsilon}\right),\cr
 &  = 2\Lambda_\tau + 3\gamma e^{2\Lambda_\tau} + {93\over 7}\gamma^2 e^{4\Lambda_\tau} + {\cal O}(\gamma^3),
 }
}
where $\epsilon$ is a UV cutoff and $L$ is an IR cutoff. From \divergences\ we find
\eqn\cutofffff{\Lambda_\tau = {1\over 2} \ln\left({L\over \epsilon}\right) - {3\over 2}\gamma {L\over \epsilon} - {15\over 7}\gamma^2 {L^2\over \epsilon^2} + {\cal O}(\gamma^3).
}
At $\gamma = 0$ we note that $\sqrt{\lambda}/ \pi\cdot 2\Lambda_\tau = (\sqrt{\lambda}/\pi)\ln(L/\epsilon)$. At the boundary we also have
\eqn\boundarycond{(\tilde\varphi, \tilde\psi) = \sqrt{2}\epsilon(\cosh(\delta/2), \pm \sinh(\delta/2)),
}
where $\delta$ is the cusp angle. Thus, we have $2\Lambda_\sigma = \delta$.

The evaluated action is
\eqn\oooooo{\eqalign{
S & = {\sqrt{\lambda}\over \pi}\left(\Lambda_\tau + 3\gamma e^{2\Lambda_\tau} + {235\over 14}\gamma^2 e^{4\Lambda_\tau}\right)\Lambda_\sigma + {\cal O}({\gamma^3}), \cr
& = \left[{\sqrt{\lambda}\over \pi}\ln\left({L \over \epsilon }\right) + {\sqrt{\lambda}\over \pi}3\gamma { L \over \epsilon} + {\sqrt{\lambda}\over \pi} {79\over 7} \gamma^2 {L^2\over \epsilon^2} + {\cal O}(\gamma^3)\right]{\delta\over 2}\cdot {1\over 2},
}
}
We first note that here in the non-conformal theory the anomalous dimension is also linear in $\delta$. We second note that at $\gamma = 0$ $S$ is proportional to $\Lambda_\tau$. We identify $\mu/\Lambda \sim L/\epsilon$. This gives the correct coefficient of the logarithmic term $f(\lambda) = {\sqrt{\lambda}/ \pi}$ \KruczenskiA. We note that we have a term linear in $\gamma$. We also note that it is negative since $-1 \leq \gamma \leq 0$. This is similar to the order $\gamma$ term in \twfff\ in the sense both are negative. The same is true for $\gamma^2$ term. In general this approach gives qualitatively (but not quantitatively) similar result to that of $E - J$ order by order in perturbation. Since the deformed boundary theory is not a CFT, we do not in general expect $E - J$ to quantitatively agree with the deformed cusp anomalous dimension \KruczenskiRT.

\newsec{Long strings as deformed sinh-Gordon solitons}

String theory on AdS spacetimes is related to integrable field theories \refs{\Pohlmeyer,\ \DeVega}. In particular the spikes of the string foldings are identified with certain soliton configurations of some integrable field theories \refs{\Jevicki,\ \JevickiK}. For example, classical string theory on $AdS_3$ is described by the sinh-Gordon model \refs{\LarsenA}. In this section we identify the deformed sinh-Gordon model that describes the spikes of the string foldings or equivalently the long strings. These are the edges of the string residing near the boundary. For large $n$, \ie, $n > n_0$, the rotating cusped strings are confined entirely near the boundary.

In this section we choose the convenient conformal gauge for the worldsheet metric. In this gauge the sigma model action is given by
\eqn\sss{S = {\sqrt{\lambda}\over 4\pi}\int d\tau d\sigma G_{AB}\partial_{\alpha}X^A\partial^{\alpha} X^B.
}
A folded rigid rotating closed string is described by the parametrization \bbbb\ with $\kappa = 0$ (which we write here again for convenience)    
\eqn\vvv{\tilde\varphi = \tau, \quad \tilde\psi = \omega \tau, \quad \tilde\theta = \tilde\theta(\sigma).
}
We have from \compon\ with $\kappa = 0$,
\eqn\aaaa{{\dot X^2} = e^{2\phi}\left(-\cosh^2\tilde\theta + \omega^2 \sinh^2 \tilde\theta\right), \quad {X'}^2 = \tilde\theta'^2, \quad {\dot X}\cdot {X'} = 0.
}
The worldsheet Virasoro constraints are 
\eqn\uuu{T_{++} = \partial_{+}X^A\partial_{+}X^B G_{AB} = 0, \quad T_{--} = \partial_{-}X^A\partial_{-}X^B G_{AB} = 0,
}
where $\sigma^+ = \tau + \sigma$ and $\sigma^- = \tau - \sigma$. Thus, it follows from the constraints 
\eqn\bbbb{\tilde\theta'^2 = e^{2\phi}\left(\cosh^2\tilde\theta - \omega^2 \sinh^2\tilde\theta\right).
}
We note that $\tilde\theta' = 0$ at $\tilde\theta = \tilde\theta_1$ where $\tilde\theta_1 = \coth^{-1}\omega$. Thus, near $\omega = 1$, $\tilde\theta_1$ approaches the boundary of ${\cal A}_3$.

We find using the dilaton \aaa\ the deformed sinh-Gordon model
\eqn\cccc{\eqalign{
\tilde\theta''  & = f(\tilde\theta; \omega, \gamma)\sinh 2\tilde\theta,\cr
& \equiv -{\partial V(\tilde\theta; \omega, \gamma)\over \partial \tilde\theta},
}
}
where $V$ gives the effective potential and the function $f$ is given by
\eqn\ccccvb{
f(\tilde\theta; \omega, \gamma)  =  {1\over 2}e^{2\phi}(1 - \omega^2) + 2\gamma e^{4\phi}\left(\cosh^2\tilde\theta - \omega^2\sinh^2\tilde\theta\right).
}

We note that in $AdS_3$, \ie, at $\gamma = 0$, the function $f$ is given by
\eqn\fadsthreee{f(\tilde\theta; \omega, 0) = {1\over 2}(1 - \omega^2).
}
We also note that at $\gamma \neq 0$ and $\omega = 1$ the function $f$ becomes 
\eqn\fadsthreeeSP{f(\tilde\theta; 1, \gamma) = 2\gamma e^{4\phi} = {2\gamma\over (1 + \gamma^2 - 2\gamma \cosh 2\tilde\theta)^2}.
}
Near $\omega = 1$, $\tilde\theta_1 = \coth^{-1}\omega$ is large and thus the cusps are near the boundary. In particular at $\gamma = -1$ we get
\eqn\fadsthreeeSPP{f(\tilde\theta; 1, -1) = -{1\over 8}{\rm sech}^4\tilde\theta.
}
For small coupling, \ie, $0 \leq -\gamma \ll 1$, we find in perturbation theory
\eqn\ccccxvv{f (\tilde\theta; \omega, \gamma) = {1\over 2}(1 - \omega^2) + \gamma \left[\left(1 - \omega^2\right)\cosh2\tilde\theta + 2\left(\cosh^2\tilde\theta - \omega^2\sinh^2\tilde\theta\right)\right] + {\cal O}(\gamma^2).
}

We rewrite the deformed sinh-Gordon equation \cccc\ using \bbbb\ as
\eqn\ccccv{\tilde\theta'' = {\tanh\tilde\theta\over  \left(1- \omega^2 \tanh^2\tilde\theta\right)}\left[(1 - \omega^2) \tilde\theta'^2 + 4\gamma\tilde\theta'^4\right].
}
Thus, we note that (at large $\tilde\theta$) the deformation corresponds to adding a quartic term in $\tilde\theta'$ with a coupling given by $\gamma$.

In this section we are in particular interested in the long strings. These are strings that extend to and/or are near the boundary. Therefore, we expand the equation \cccc\ around $\omega = 1$. With $\omega = 1 + 2\delta$ where $0 \leq \delta \ll 1$, the equation \cccc\ becomes
\eqn\ccccv{\eqalign{
\tilde\theta'' & = 2\left[\gamma - \delta(1 - \gamma)^2\right]e^{4\phi}\sinh 2\tilde\theta,\cr
& = {2\gamma \sinh2\tilde\theta\over (1 + \gamma^2 - 2\gamma \cosh 2\tilde\theta)^2} + {\cal O}(\delta),
}
}
which we rewrite using \bbbb\ and $\delta = 0$ as
\eqn\ccccvv{\tilde\theta'' - 2\gamma \tilde\theta'^4 \sinh 2\tilde\theta = 0.
}
The effective potential $V$ at $\delta = 0$ is given by
\eqn\potentialNEW{V = -{1\over 2}\cdot {1\over 1 + \gamma^2 - 2\gamma \cosh 2\tilde\theta} + V_0 = -{1\over 2}e^{2\phi} + V_0.
}
where $V_0$ is an integration constant which in general depends on the coupling $\gamma$. It gives the asymptotic value of the potential.

In the special case where $\gamma = -1$ the equation \ccccv\ reduces to
\eqn\ccccvvz{\tilde\theta'' + {1\over 4}\tanh\tilde\theta{\rm sech}^2\tilde\theta = 0.
}
In the large $\tilde\theta$ limit this gives
\eqn\largeccccvvz{\tilde\theta'' + e^{-2\tilde\theta} = 0.
}
In the case $\gamma = 0$ and $\tilde\theta$ is large the sinh-Gordon model \cccc\ with $f(\tilde\theta; \omega, 0)$ given in \fadsthreee\ reduces to
\eqn\ddddxxu{
\tilde\theta'' + \delta \cdot e^{2\tilde\theta} = 0.
}
This is the Liouville equation.

An interesting limit is the (Poincar\'e) limit where $0 \leq -\gamma \ll -1$ and $\tilde\theta \gg 1$ with $\gamma e^{2\tilde\theta}$ fixed and finite. It follows from \cccc\ that 
\eqn\ccccvvzxz{\tilde\theta'' - {1\over 2}\gamma F = 0, \quad k\alpha' F = H_{012},
} 
where $H_{012}$ is the three form flux \bbbbbbv\ (written here agian for convenience)
\eqn\ccccvfluxvzxz{ H_{012} = {2k\alpha' e^{2\tilde\theta}\over \left(1 - \gamma e^{2\tilde\theta}\right)^2}.
} 
The effective potential \potentialNEW\ now takes the form
\eqn\potlimm{V = -{1\over 2}{1\over (1 - \gamma e^{2\tilde\theta})} + V_0.
}
 We note that $F = -(2/\gamma)\partial V/\partial\tilde\theta$ which is also valid at finite values of the coupling $\gamma$.
 
In perturbation \ccccxvv\ or \ccccvv\ gives to leading order in the coupling $\gamma$
\eqn\ccccvvvv{\tilde\theta'' - 2\gamma  \sinh 2\tilde\theta = 0.
}
In the limit where $\tilde\theta$ is large but small $\gamma e^{2\tilde\theta}$, we get
\eqn\largeccccvvvv{\tilde\theta'' - \gamma  e^{2\tilde\theta} = 0.
}
see also \ccccvvzxz\ or \potlimm. Thus, at weak or small coupling this is similar to the Liouville equation \ddddxxu\ except that $\delta$ is replaced by $-\gamma$. We leave a detailed and thorough discussion of the modified sinh-Gordon model \cccc\ for a future work.

\newsec{Discussion}

In this paper we studied the structure of large angular momentum expansion of the quantity $E - J$ for certain rotating closed strings in an asymptotically non-AdS spacetime denoted ${\cal A}_3$ \aaa, where $E$ is the energy and $J$ is the angular momentum of the spinning strings. The spacetime ${\cal A}_3$ \aaa\ interpolates between $AdS_3$ and (asymptotically) $\IR \times S^1 \times \IR$ linear dilaton spacetime (times a compact internal manifold ${\cal M}_7$). In \AsratMHB\ the deformation is interpreted as moving the conformal boundary of $AdS_3$. This breaks the spacetime conformal symmetry.

In $AdS_3$ the quantity $E - J$ gives the anomalous dimensions of certain twist two and higher operators of the dual undeformed CFT. In the context of AdS/CFT correspondence, twist two and higher operators are described by certain rotating folded rodlike \Gubser\ and cusped closed strings \KruczenskiA, respectively.

We showed that the quantity $E - J$ in general is bounded. For a folded rodlike rotating closed string it is bounded by \twtomegaone
\eqn\twtomegaoneET{E - J = {2\sqrt{\lambda}\over \pi(1- \gamma)}K\left(\left({1 + \gamma\over 1 - \gamma}\right)^2\right).
}
The energy and angular momentum of the spinning string are not bounded. Thus, above a certain energy the closed strings will be unstable and fragment into segments. In perturbation we found in the large angular momentum limit, \ie, $J\gg \sqrt{\lambda}$, the result \twfff
\eqn\twisttwoanomalous{
E - J = {\sqrt{\lambda}\over \pi}\left[\ln\left(2\pi {J\over \sqrt{\lambda}}\right) + {2\pi \over 3}{J\over \sqrt{\lambda}}\gamma + {\cal O}\left({J^2\over \lambda}\gamma^2\right)\right].
}
The first term is the anomalous dimension of a certain twist two operator in the undeformed theory in the large spin $S = J$ limit. We also note $E - J$ do not scale linearly with $n$, see \refs{\Callan,\  \Belitsky} for a related discussion. In the approach based on a Wilson loop (containing a single cusp singularity) we found qualitatively similar result \oooooo\ for the cusp anomalous dimension. Thus, in the large $J$ limit, we may view $E - J$ \twisttwoanomalous\ as the (deformed) anomalous dimension of certain (twist two or higher) operator with large spin along the deformation or trajectory parametrized by the coupling $\gamma$. We hope to clarify and verify this in a future work.

In the case $\gamma = -1$ we obtained the simple exact Regge relation \kkkkkkkkkk,
\eqn\twtlomegagminusRegg{E^2 = J(J + \lambda^{1\over 2}).
}

For short strings we found \twtlomega
\eqn\twtlomegaXT{
E = \sqrt{{2\over 1 - \gamma}}\lambda^{1\over 4}J^{1\over 2} + {\cal O}(\lambda^{3/4}J^{3/2}).
}
Note in the short string case the power of $\lambda$ is $1/4$.

In the case of rotating closed strings with $n$ spikes or cusps we have two distinct cases depending on whether $n$ is greater or less than $n_0$. The minimum value $n_0 = \lceil{\mu} \rceil$ where 
\eqn\muagain{
\mu := (1 - \gamma)\sqrt{1 + \gamma^2\over 2\gamma^2}, \quad -1 \leq \gamma \leq 0,
}
and $\lceil{\mu}\rceil$ gives the integer greater than or equal to $\mu$. The variable $\mu$ satisfies $\mu(1/\gamma) = \mu(\gamma)$. Thus, $\gamma = -1$ is a self dual point. At $\gamma = -1$ we have $\mu(-1) = 2$. At $\gamma = 0$ $n$ denotes the twist of a certain operator in the original undeformed theory and the quantity $E - J$ gives its anomalous dimension.

For $n \geq n_0$ the quantity $E - J$ is bounded by \twhgggg
\eqn\twhggggagain{E -  J =  {\sqrt{\lambda}\over 4}\cdot {(1 - \gamma) \over - \gamma }.
}
For $n \leq n_0$ it is bounded by \maxefbe
\eqn\maxefbeagain{
E -  J  = {n\sqrt{\lambda}\over \pi}{\sqrt{(1 - \gamma)^2 - 2\gamma (x_0 - 1)}\over \sqrt{(1 - \gamma)^4 - 4\gamma(1+\gamma^2)(x_0 - 1)}}K\left({(1 - \gamma)^2(1 + \gamma)^2}\over (1 - \gamma )^4 - 4\gamma(1 + \gamma^2)(x_0 - 1)\right),
}
where $x_0 = \cosh 2\tilde\theta_0$ is determined using the angle difference constraint \integralthetaiiii. In perturbation $E - J$ is given in the large $J$ limit by \correctiondi
\eqn\correctionanomalus{E - J = {n\sqrt{\lambda}\over 2\pi}\left[\ln\left({4 \pi \over n} {J\over \sqrt{\lambda}}\right) + {4\pi \over 3n}{J\over \sqrt{\lambda}}\gamma +  {\cal O}(\gamma^2)\right], \quad n \geq 2. 
}
The first term is the anomalous dimension of a twist $n$ operator in the undeformed theory.

We also showed that for $n > n_0$ both $E$ and $J$ are bounded. The maximum angular momentum is given by \maxjjj
\eqn\maxjjjagain{J = {\sqrt{\lambda}\over 4(1 - \gamma)}\left({n\mu^2\over n - \mu}\right), 
}
where $\mu$ is given above in \muagain.

For $\gamma = -1$ we found the simple exact Regge relation \specRegge
\eqn\specReggeagain{m^2 = 4\left({n - 1\over n^2}\right)\cdot (l + n)l, \quad n \geq  2,
}
where $m$ and $l$ are given in \renormal
\eqn\renormalagain{
m:= {2E\over \sqrt{\lambda}}, \quad l:= {2J\over \sqrt{\lambda}}.
}
Since $n_0 = 2$ for $\gamma = -1$, both $E$ and $J$ are bounded for $n > n_0 = 2$; see Fig. 4.

We also obtained the deformed sinh-Gordon equation that describes the long strings that extend to the boundary. It is given by \cccc
\eqn\ccccagain{\eqalign{
\tilde\theta''   & = {1\over 2}\gamma F, \quad F = {H_{012}\over k\alpha'},
\cr
& = -{\partial V(\tilde\theta; \gamma)\over \partial \tilde\theta} },
}
where $H_{012}$ is the three form flux \bbb\ and $V$ is the effective potential and it is given by \potentialNEW
\eqn\potentialNEWagain{V(\tilde\theta; \gamma) = -{1\over 2}\cdot {1\over 1 + \gamma^2 - 2\gamma \cosh 2\tilde\theta} + V_0(\gamma),
}
where $V_0$ is the asymptotic value of the potential. We hope to study further the deformed sinh-Gordon equation in a future work. We hope to construct an effective field theory that reproduces \ccccagain\ as an equation of motion. This will fully address the question raised in \Meseret\ concerning the single trace $T\bar T$ deformation in non-orbifold theories.

In the long string sector and at weak coupling, string theory on the background \aaa\ is (believed) to be described by a $T{\bar T}$ deformed symmetric product theory; see \AsratMHB\ and references therein for a detailed discussion. We hope to identify the precise form of the twist two and higher operators that are described in the dual string theory on $AdS_3$ by folded rodlike and spiky closed strings in a future work. We also hope to compute their deformed anomalous dimensions and discuss their precise relations with the quantity $E - J$ \correctionanomalus\ and ${\sqrt{\lambda}\over 2\pi}\cdot 2n\tilde\theta_1$ (see \lenn) \Braga. In particular we consider the operators with large spin since these are the ones described on ${\cal A}_3$ by long strings.

\bigskip\bigskip
\noindent{\bf Acknowledgements:} This work is supported by the Department of Atomic Energy under project no. RTI4001. I thank Ofer Aharony for valuable comments on an earlier draft of the paper.

\appendix{A}{Intermediate results}

We collect the intermediate results required in section two.

\eqn\intone{\eqalign{
I_1& = \int_{x_0}^{x_1} dx \sqrt{(x-x_0)(x + x_0 + 1)\over x_1 - x},\cr
& = {2\over 3}\sqrt{x_1 + x_0 + 1}\left[(2x_1 + 1)E\left({x_1 - x_0\over x_1 + x_0 + 1}\right) - (2x_0 + 1)K\left({x_1 - x_0\over x_1 + x_0 + 1}\right)\right],\cr
}
}
where $E(\alpha)$ is the complete elliptic integral of the second kind
\eqn\twhhhc{E(\alpha) = \int_0^{\pi\over 2}\sqrt{1 - \alpha \sin^2\theta}d\theta,
}
and $K(\alpha)$ is the complete elliptic integral of the first kind,
\eqn\kkkkkkkk{K(\alpha) = \int_{0}^{{\pi\over 2}}{d\theta\over \sqrt{1 - \alpha \sin^2\theta}}.
}

\eqn\inttwo{\eqalign{
I_2& = \int_{x_0}^{x_1}{dx\over x + 1}\sqrt{(x- x_0)(x + x_0 + 1)\over (x_1 - x)},\cr
& =  2\left[\sqrt{x_1 + x_0 + 1}E\left({x_1 - x_0\over x_1 + x_0 + 1}\right)-\right. \cr
&  {x_0 + 1\over \sqrt{x_1 + x_0 + 1}}K\left({x_1 - x_0\over x_1 + x_0 + 1}\right) - \cr
&\left.{x_0(x_0 + 1)\over (x_1 + 1)\sqrt{x_1 + x_0 + 1}}\Pi\left({x_1 - x_0\over x_1 + 1},{x_1 - x_0\over x_1 + x_0 + 1}\right)\right],\cr
}
}
where $\Pi(\alpha, \beta)$ is the complete elliptic integral of the third kind,
\eqn\kkkkkkk{\Pi(\alpha, \beta) = \int_{0}^{{\pi\over 2}}{d\theta\over (1 - \alpha \sin^2\theta)\sqrt{1 - \beta \sin^2\theta}}.
}

\eqn\intthree{\eqalign{
I_3& = \int_{x_0}^{x_1} dx x\sqrt{(x-x_0)\over (x_1 - x)(x + x_0 + 1)},\cr
& = {2\over 3}\left[\sqrt{x_1 + x_0 + 1}\left[2(x_1 - x_0 - 1) - x_0\right]E\left({x_1 - x_0\over x_1 + x_0 + 1}\right) - \right.\cr
&\left.  {2x_0 + 1\over \sqrt{x_1 + x_0 + 1}}\left[2(x_1 - x_0 - 1) - x_1\right]K\left({x_1 - x_0\over x_1 + x_0 + 1}\right)\right].
}
}

\eqn\intfour{\eqalign{
I_4 & = \int_{x_0}^{x_1}dx \sqrt{x_1 - x\over (x - x_0)(x + x_0 + 1)},\cr
& = 2\sqrt{x_1 + x_0 + 1}\left[K\left({x_1 - x_0\over x_1 + x_0 + 1}\right) - E\left({x_1 - x_0\over x_1 + x_0 + 1}\right)\right].\cr
}
}
\eqn\intfive{\eqalign{
I_5  & = \int_{x_0}^{x_1}dx (2x + 1)\sqrt{x_1 - x\over (x - x_0)(x + x_0 + 1)},\cr
& = {2\over 3}\sqrt{x_1 + x_0 + 1}\left[(1 + 2x_1)E\left({x_1 - x_0\over x_1 + x_0 + 1}\right) - (1 + 2x_0)K\left({x_1 - x_0\over x_1 + x_0 + 1}\right)\right].\cr
}
}
\eqn\intsix{\eqalign{
I_6  & = \int_{x_0}^{x_1}dx {2x(x + 1)\over x + x_0 + 1}\sqrt{x_1 - x\over (x - x_0)(x + x_0 + 1)},\cr
& = {4\over 3}\sqrt{x_1 + x_0 + 1}\left[C_0E\left({x_1 - x_0\over x_1 + x_0 + 1}\right) -C_1 K\left({x_1 - x_0\over x_1 + x_0 + 1}\right)\right],
}
}
where 
\eqn\constE{C_0 = {2+x_1 + 10 x_0 + 2x_0 x_1 + 9 x_0^2\over 2x_0 + 1}, \quad C_1 = {2 + 9x_0 + 7x_0^2 + 2x_1 + 4 x_0 x_1\over 1 + x_1 + x_0}.
}

\listrefs
\end